\documentclass[lettersize,journal]{IEEEtran}
\usepackage{amsmath,amsfonts}
\usepackage{algorithmic}
\usepackage{algorithm}
\usepackage{array}
\usepackage[caption=false,font=normalsize,labelfont=sf,textfont=sf]{subfig}
\usepackage{textcomp}
\usepackage{stfloats}
\usepackage{url}
\usepackage{tcolorbox}
\usepackage{verbatim}
\usepackage{graphicx}
\usepackage{cite}
\usepackage{hyperref}
\usepackage{xcolor}
\usepackage{threeparttable}
\usepackage{booktabs}

\newtcolorbox{boxA}{
    fontupper = \bf,
    boxrule = 1.5pt,
    colframe = black 
}

\mathchardef\mhyphen="2D

\newcommand{\ClientAction}[1]{ 
	\node[right] at (\InitX, \Y) {#1};
}
\newcommand{\ServerAction}[1]{
	\node[left] at (\RespX, \Y) {#1};
}

\newcommand{\ClientToServer}[3][->]{
	\NextLine[0.5]
	\draw[#1] (\ArrowLeft,\Y) -- node[above] {#2} node[below] {#3} (\ArrowRight,\Y) ;
	\NextLine[0.5]
}
\newcommand{\ServerToClient}[3][->]{
	\NextLine[0.5]
	\draw[#1] (\ArrowRight,\Y) -- node[above] {#2} node[below] {#3} (\ArrowLeft,\Y) ;
	\NextLine[0.5]
}

\newcommand{\NextLine}[1][1.0]{
	\pgfmathparse{\Y+#1}
	\edef\Y{\pgfmathresult}
}
\newcommand{\linkgame}[2]{\hyperref[#1]{G#2}}

\newcounter{Bdversary}

\newcommand\orcidicon[1]{\href{https://orcid.org/#1}{\mbox{\scalerel*{
\begin{tikzpicture}[yscale=-1,transform shape]
\pic{orcidlogo};
\end{tikzpicture}
}{|}}}}

\newcommand*{\addFileDependency}[1]{
  \typeout{(#1)}

  \IfFileExists{#1}{}{\typeout{No file #1.}}
}

\newif\iffullversion
\newif\iffullversion
\newif\ifsubmissionversion
\fullversiontrue
\submissionversionfalse

\usepackage{adjustbox}
\usepackage{tikz}

\tikzset{
  orcidlogo/.pic={
    \fill[orcidlogocol] svg{M256,128c0,70.7-57.3,128-128,128C57.3,256,0,198.7,0,128C0,57.3,57.3,0,128,0C198.7,0,256,57.3,256,128z};
    \fill[white] svg{M86.3,186.2H70.9V79.1h15.4v48.4V186.2z}
                 svg{M108.9,79.1h41.6c39.6,0,57,28.3,57,53.6c0,27.5-21.5,53.6-56.8,53.6h-41.8V79.1z M124.3,172.4h24.5c34.9,0,42.9-26.5,42.9-39.7c0-21.5-13.7-39.7-43.7-39.7h-23.7V172.4z}
                 svg{M88.7,56.8c0,5.5-4.5,10.1-10.1,10.1c-5.6,0-10.1-4.6-10.1-10.1c0-5.6,4.5-10.1,10.1-10.1C84.2,46.7,88.7,51.3,88.7,56.8z};
  }
}

\usetikzlibrary{svg.path}
\usetikzlibrary{calc}

\begin{document}


\title{Leveraging A New GAN-based Transformer \\ with ECDH Crypto-system for Enhancing \\Energy Theft Detection in Smart Grid}

\author{Yang Yang,~\IEEEmembership{Student Member,~IEEE,} Xun Yuan, Arwa Alromih,
Aryan Mohammadi Pasikhani,~\IEEEmembership{Member,~IEEE,}\\
Prosanta Gope,~\IEEEmembership{Senior Member,~IEEE,}
Biplab Sikdar,~\IEEEmembership{Senior Member,~IEEE,}}



\markboth{IEEE Transactions on Dependable and Secure Computing}%
{Shell \MakeLowercase{\textit{et al.}}:IEEE Transactions on Dependable and Secure Computing}



\maketitle


\begin{abstract}

Detecting energy theft is vital for effectively managing power grids, as it ensures precise billing and prevents financial losses. Split-learning emerges as a promising decentralized machine learning technique for identifying energy theft while preserving user data confidentiality. Nevertheless, traditional split learning approaches are vulnerable to privacy leakage attacks, which significantly threaten data confidentiality.  To address this challenge, we propose a novel GAN-Transformer-based split learning framework in this paper. 
This framework leverages the strengths of the transformer architecture, which is known for its capability to process long-range dependencies in energy consumption data. Thus, it enhances the accuracy of energy theft detection without compromising user privacy.
A distinctive feature of our approach is the deployment of a novel mask-based method, marking a first in its field to effectively combat privacy leakage in split learning scenarios targeted at AI-enabled adversaries. This method protects sensitive information during the model's training phase. Our experimental evaluations indicate that the proposed framework not only achieves accuracy levels comparable to conventional methods but also significantly enhances privacy protection. The results underscore the potential of the GAN-Transformer split learning framework as an effective and secure tool in the domain of energy theft detection.

\end{abstract}


\section{Introduction}
\IEEEPARstart{T}he advent of Smart Grids (SG) marks a pivotal shift in the evolution of smart cities, reshaping the energy distribution landscape by integrating advanced digital technologies and many sensors into the conventional grid infrastructure. This modernization has given rise to a dynamic and interactive energy network equipped with capabilities for real-time data analytics, bidirectional communication systems, and decentralized energy management. These advancements in smart grids have opened doors to remarkable improvements in energy efficiency, substantial reduction in environmental impacts, and bolstered resilience of the grid against various contingencies. However, integrating smart technologies has also introduced significant cybersecurity threats. These concerns are most pronounced in areas such as energy theft \cite{alromih2022privacy, bbc}, which refers to unauthorized and illicit interference with the measurement of electricity consumption or generation, thus bypassing established billing protocols \color{blue}{(discussed in section \ref{sec:adversary_model}) }. \color{black}{Furthermore}, the advent of these technologies has heightened the risk of privacy leaks \cite{grid2010guidelines}, where confidential information is susceptible to unauthorized exposure or exploitation.
This prevalent issue not only undermines the financial sustainability of power grids but also unjustly burdens legitimate consumers with escalated costs \cite{smith2004electricity}. In response, utility companies have leveraged cutting-edge, privacy-preserving energy theft detection systems. These systems are engineered to maintain a delicate equilibrium between detecting and mitigating energy theft incidents and upholding the confidentiality of consumer data. They incorporate sophisticated cryptographic methods, secure data-sharing frameworks, and advanced privacy-enhancing technologies \cite{wu2022p2detect}. The aim is to anonymize consumer identities and specific usage patterns while meticulously analyzing energy consumption data for signs of fraudulent activities.
 Despite the potential of privacy-preserving energy theft detection systems, they are not without challenges. This paper delves into these challenges \cite{5054916}, explicitly focusing on cybersecurity aspects. We will dissect the vulnerabilities and potential attack vectors that threaten these systems' integrity and functionality. The discussion will encompass the technical nuances of energy theft detection, scrutinize various privacy preservation techniques, identify their inherent weaknesses susceptible to exploitation by adversaries, and explore the complex adversarial environment that must be navigated to fortify both the security of the smart grid and the privacy of its consumers.

\subsection{Related Work and Motivation}
Several studies in the literature have proposed models for detecting energy theft while preserving privacy.  These studies employ a variety of privacy-related techniques, including data encryption, data anonymity, federated learning and split learning. Notably, the paper \cite{wu2022p2detect} introduced \emph{p2Detect} a model that utilizes homomorphic encryption for detection. This approach enables the model to process encrypted data, thus eliminating the need to use data in its unencrypted form.
Another encryption-based technique is used in \cite{ibrahem2022privacy}. The authors introduce a novel solution that uses a functional encryption cryptosystem and a decentralized aggregation scheme. This solution eliminates the need for a central key distribution centre by allowing the detection stations to securely send encrypted training parameters to an aggregator without exposing sensitive information. In \cite{zhao2023privacy}, the authors presented a privacy-preserving electricity theft detection scheme based on blockchain technology, eliminating the need for a third party. This scheme employs an enhanced functional encryption system to enable theft detection and load monitoring while safeguarding consumers' privacy. Additionally, we utilize distributed blockchain storage for consumers' data to mitigate concerns related to data tampering and other security threats.
Research on distributed machine learning-based energy theft detection includes the use of federated learning and split learning. The paper in \cite{ashraf2022feddp} proposed \emph{FedDP}, which is a novel Federated Voting Classifier (FVC) for accurate energy theft identification. FVC combines the results of several traditional machine learning classifiers to enhance detection accuracy. Along with functional encryption, Federated learning was also used in \cite{ibrahem2022privacy}. Moreover, \cite{wen2021feddetect} used federated learning to protect the privacy of customers' data. The other type of distributed ML-based approach is split learning, which was used as an energy theft detector in \cite{alromih2022privacy}. The authors proposed an enhanced version of split learning, enabling it to be directly applied in the smart grid (SG) environment. Moreover, the paper claims that splitting the detection model makes the system more robust against honest but curious adversaries. 

\textbf{Problem Statement and Motivation:}

In the field of privacy-preserving power theft detection, high accuracy rates are often cited as a key strength of existing methodologies. However, a deeper examination reveals significant shortcomings. For instance, encryption-based methods face considerable communication and computational overheads \cite{muzumdar2022designing}. Moreover, these methods' reliance on key distribution centres introduces a vulnerability due to the risk of a single point of failure, posing a critical cybersecurity concern.
The landscape is further complicated by inherent drawbacks in privacy-preserving machine learning techniques. Notably, federated learning frameworks, while innovative, suffer from high communication costs due to frequent interactions between the central server and its client nodes \cite{gosselin2022privacy, briggs2021review}. Additionally, the challenge of managing non-IID data distributions among these clients hampers the effective aggregation of a global model, undermining the overall efficacy of federated learning systems.

The advent of split learning was initially seen as a solution for some of these challenges faced by federated learning. However, subsequent research revealed that split learning is prone to various privacy-related attacks, an issue not initially considered in its development \cite{rigaki2020survey}. This emerging field of research concentrates on attacks capable of compromising information about the machine learning model or its data, including reconstruction, membership inference, property inference, and model extraction attacks. For example, during reconstruction or inference attacks, the data communicated in the training process can be leveraged to deduce sensitive information about the input data \cite{salem2020updates}. These vulnerabilities persist even when models are implemented with privacy-preserving techniques, indicating the necessity for additional protective measures.

Moreover, within the context of machine learning models utilising federated learning or other privacy-centric methodologies, there exists the risk of membership inference attacks. These attacks exploit subtle discrepancies in model outputs to determine whether a specific data point was included in the training dataset \cite{shokri2017membership}. Such attacks can lead to inadvertent information leakage concerning individual data points. In conclusion, while current privacy-preserving approaches in power theft detection primarily focus on data privacy, they often overlook the broader spectrum of privacy vulnerabilities inherent in these methodologies. There is a critical need to address these gaps to ensure comprehensive protection against the multifaceted threats faced in the realm of smart grid cybersecurity.

\subsection{Contributions}

In this research, we introduce an \emph{innovative} approach using a new variant of Generative Adversarial Network (GAN)-based transformers for detecting electrical energy theft, integrating advanced split-learning techniques to safeguard user data privacy. The unique structure of GAN models presents challenges in their straightforward application to split learning. To overcome this, we innovatively combine GAN with split learning, balancing user data protection and a marginal compromise in model accuracy. Our approach includes the development of diverse split-learning frameworks tailored for GANs, catering to both rapid training and privacy preservation. The performance analysis demonstrates that the performance of our proposed scheme is significantly better than any state-of-the-art schemes in energy theft detection (as depicted in Table \ref{tab:auc}) and Table \ref{tab:r2}). 
Embarking on new frontiers in smart grid security and data privacy, this study introduces several groundbreaking advancements in the realm of electrical energy theft detection. We present the following key contributions:

\begin{itemize}
    \item \textbf{A new variant of GAN-Based Transformer for Smart Grids:} We propose a cutting-edge GAN-based transformer model uniquely designed for energy theft detection in smart grids. To the best of our knowledge, this model is the first of its kind, showcasing the effective integration of transformer and GAN-based adversary loss in tackling the issue of energy theft.
    
    \item \textbf{Modeling Framework for GAN and Split Learning Integration:} Our work is \emph{the first} to present a protocol-level modelling framework that synthesizes GAN with split learning. Given the distinct architecture of GAN models, we have crafted a highly suitable GAN-based segmentation learning model, representing a significant leap in this area of research.

    \item \textbf{Protocol-Level Defense Mechanism:} To enhance the security of split learning against AI-enabled reconstruction attacks \cite{salem2020updates}, we have devised a robust protocol-level defence strategy. This novel integration of machine learning and cryptography significantly increases the resilience of our model against such sophisticated cyber threats while maintaining a higher level of efficiency.

    \item \textbf{Comprehensive Model Evaluation with Smart Grid Dataset:} Through comprehensive comparative analyses with the state-of-the-art (SOTA) models, our proposed model emerges as a top performer, showcasing exceptional performance and broad utility across various adversary levels. Our model's performance is convincingly reinforced by conducting thorough assessments using the Pecan Street smart grid dataset \cite{pecanstreet-dataport-2023}\footnote{\textbf{Our research leverages the Pecan Street dataset, sourced from the comprehensive Dataport repository of Pecan Street Inc. This dataset represents a pivotal resource in our model evaluation, offering detailed, circuit-level electricity use data at one-minute to one-second intervals for approximately 1000 homes in the United States, with PV generation and EV charging data for a subset of these homes. For our research purposes, we obtained a paid license from Pecan Street, ensuring full compliance with their data usage policies and restrictions.}} (results shown in table \ref{tab:auc}). 

\end{itemize}

These contributions collectively mark a significant advancement in the realm of smart grid security, particularly in addressing the challenges of energy theft detection, while concurrently ensuring stringent user privacy protection.

\begin{table}[ht]
\caption{Notions and Cryptographic Functions}
\centering
\begin{tabular}{l | l }

\toprule[1.5pt]

\textbf{Symbols} & \makebox[0.30\textwidth] {\textbf{Description}}\\

\toprule[1.5pt]

$ \theta $ & Model parameter  \\
$Enc,\ Dec$ & Encoder and decoder of the generator  \\
$Dis$ &  Discriminator  \\
$(PK,\ SK)$ & Public and secret key pair  \\
$G, d, Q$ & Parameter for ECDH  \\
$k_{Enc},\ k_{Mask}$ & Shared key for encryption and mask  \\
$\Phi$ & pseudorandom generator  \\
$T_{Mid}$ & Intermediate data  \\
$NN$ & Neural Network  \\

\toprule[1.5pt]
\end{tabular}
\label{tab:notions}
\vspace{-6mm}  

\end{table}

\section{Preliminaries}

This section introduces the preliminaries of the Generative Adversarial Network and Transformer encoder, which are applied in the proposed GAN-Transformer.

\subsection{Generative Adversarial Network}

The Generative Adversarial Network (GAN) \cite{goodfellow2014generative} aims to generate data following a distribution similar to the training data from a fixed and simple distribution, e.g. Gaussian distribution. This goal is achieved by iteratively training a discriminator and a generator. The discriminator is trained to distinguish between the real samples and the synthetic samples generated by the generator. The generator is trained to fool the discriminator by producing better synthetic samples. The above objectives are achieved by optimizing the following min-max problem,
\begin{equation}
\begin{aligned}
\label{eq:Xun gan objectives}
\min _G \max _D V(D, G)= &\mathbb{E}_{\boldsymbol{x} \sim p_{\text {data }}(\boldsymbol{x})}[\log D(\boldsymbol{x})]+ \\
                        &\mathbb{E}_{\boldsymbol{z} \sim p_{\boldsymbol{z}}(\boldsymbol{z})}[\log (1-D(G(\boldsymbol{z})))],
\end{aligned}
\end{equation}
where $D$ denotes the discriminator, $G$ denotes the generator, $p_{\text{data}}(x)$ denotes the distribution of the real data, and $p_{\text{z}}(z)$ denotes the fixed simple distribution.

On the other hand, GAN has been used for anomaly detection, e.g. GANomaly \cite{akcay2018ganomaly}. Different from the standard GAN, the generator of GANomaly is trained to reconstruct the input real samples and the discriminator is trained to distinguish between the real samples and the reconstructed samples. Unlike the vanilla GAN where $G$ is updated based on the output of $D(\text{real/fake})$, $G$ is updated based on the internal representation of $D$ in GANomaly. Formally, let $f$ be a function that outputs an intermediate layer of the discriminator. For a given input x drawn from the input data distribution $p_{\text {data }}(\boldsymbol{x})$, the objective is to minimize the L2 distance between the feature representation of the original and the corresponding generated samples. Hence, the adversarial loss $\mathcal{L}_{a d v}$ is calculated as:
\begin{equation}
\label{eq:Xun Ganomaly}
\mathcal{L}_{a d v}=\mathbb{E}_{x \sim p_{\text {data }}(\boldsymbol{x})} \| f(x)-\mathbb{E}_{x \sim p_{\text {data }}(\boldsymbol{x})} f\left(G(x)\right) \|_2 .
\end{equation}

\begin{figure*}[htb]
\center{\includegraphics[scale=0.40,trim=0 0 0 0,clip]{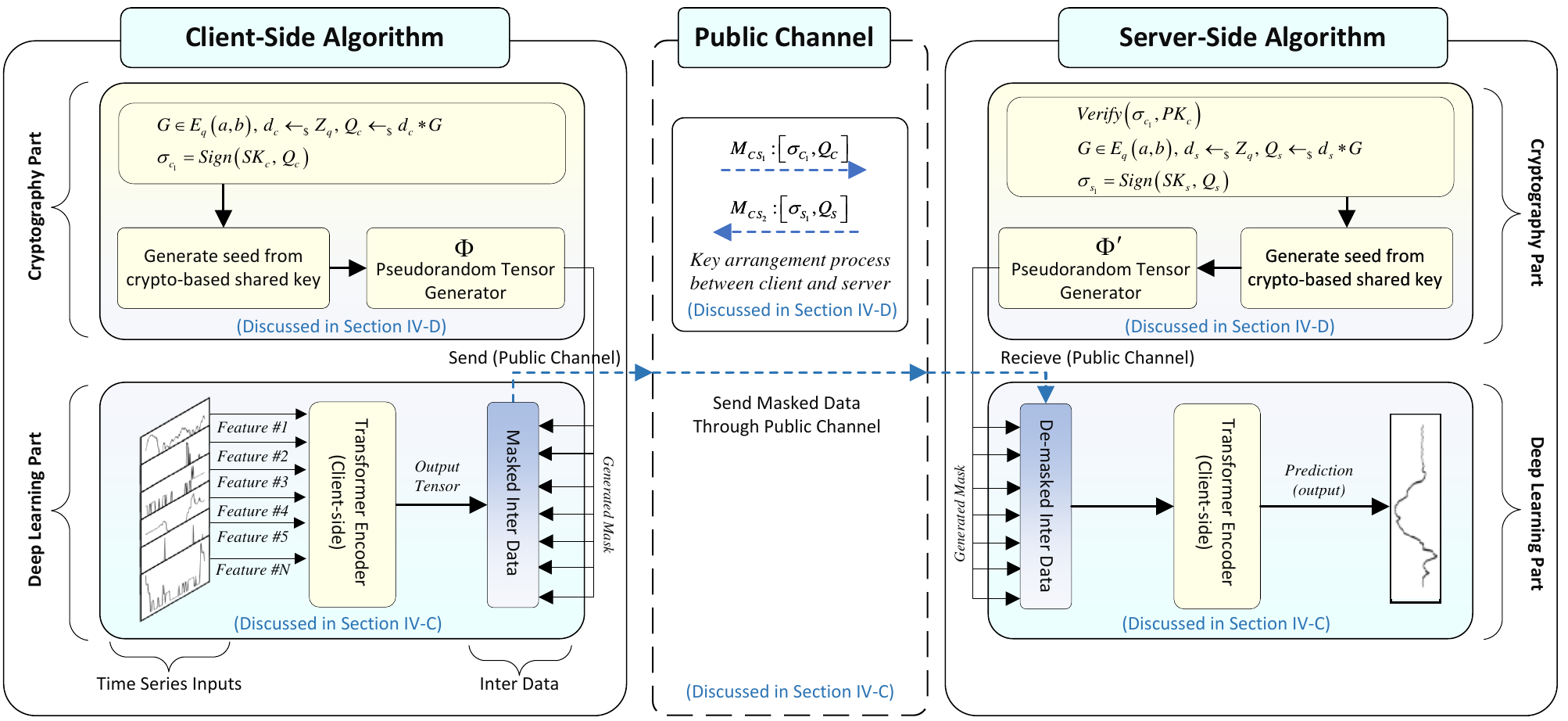}}
\caption{{System Model}}
\label{fig:sys_model} 
\vspace{-6mm}
\end{figure*}

\subsection{Transformer Encoder}

Transformer \cite{vaswani2017attention} was initially proposed for language understanding tasks. Recently, the powerful feature learning ability of Transformer was discovered, which has been taken advantage of for computer vision tasks. The most important mechanism of Transformer is the multiheaded self-attention (MSA) mechanism. First of all, the self-attention (SA) mechanism can be expressed by the followings,
\begin{equation}
\label{eq:Xun SA}
\begin{aligned}
{[\mathbf{q}, \mathbf{k}, \mathbf{v}] } & =\mathbf{z} \mathbf{U}_{q k v} & \mathbf{U}_{q k v} & \in \mathbb{R}^{D \times 3 D_h} \\
A & =\operatorname{softmax}\left(\mathbf{q} \mathbf{k}^{\top} / \sqrt{D_h}\right) & A & \in \mathbb{R}^{N \times N} \\
\mathrm{SA}(\mathbf{z}) & =A \mathbf{v}, & &
\end{aligned}
\end{equation}
where $z$ is the input of the SA mechanism and $\mathbf{U}_{q k v}$ denotes the parameters of the SA mechanism. Then, the MSA mechanism can be expressed as,
\begin{equation}
\label{eq:Xun MSA}
\operatorname{MSA}(\mathbf{z})=\left[\mathrm{SA}_1(z) ; \mathrm{SA}_2(z) ; \cdots ; \mathrm{SA}_k(z)\right] \mathbf{U}_{m s a},
\end{equation}
where $\mathbf{U}_{m s a} \in \mathbb{R}^{k \cdot D_h \times D}$ denotes the parameters of the MSA mechanism.

The Transformer encoder consists of alternating layers of MSA and MLP blocks. Layernorm (LN) is applied before every block and residual connections are applied after every block. The Transformer encoder can be expressed as the following equations,
\begin{equation}
\label{eq:Xun Transformer encoder}
\begin{aligned}
\mathbf{z}_0 & =\left[\mathbf{x}_{\text {class }} ; \mathbf{x}_p^1 \mathbf{E} ; \mathbf{x}_p^2 \mathbf{E} ; \cdots ; \mathbf{x}_p^N \mathbf{E}\right]+\mathbf{E}_{p o s},\\
\mathbf{z}_{\ell}^{\prime} & =\operatorname{MSA}\left(\operatorname{LN}\left(\mathbf{z}_{\ell-1}\right)\right)+\mathbf{z}_{\ell-1}, & & \ell=1 \ldots L \\
\mathbf{z}_{\ell} & =\operatorname{MLP}\left(\operatorname{LN}\left(\mathbf{z}_{\ell}^{\prime}\right)\right)+\mathbf{z}_{\ell}^{\prime}, & & \ell=1 \ldots L \\
\mathbf{y} & =\operatorname{LN}\left(\mathbf{z}_L^0\right), & &
\end{aligned}
\end{equation}
where $\mathbf{x}_{\text {class }}$ is the label, $\mathbf{x}_p=[\mathbf{x}_p^1, \cdots, \mathbf{x}_p^N]$ is the input vector, $\mathbf{E}$ and $\mathbf{E}_{p o s}$ are pre-trained embedding matrices, and $y$ is the output of Transformer encoder.

\subsection{Elliptic Curve Diffie-Hellman}
The Elliptic Curve Diffie-Hellman (\textbf{ECDH}) \cite{ecdh} is an elliptic curve cryptography based key exchange protocol. It facilitates secure key exchange between two parties by utilizing elliptic curves:

\begin{equation}
    y^2 = x^3 + ax + b
\end{equation}

where $a$ and $b$ are constants. The ECDH protocol enables the establishment of a shared key through an insecure communication channel by leveraging the mathematical properties of eliptic curves. Two algorithms defined by ECDH are Key Gen and Key arrangement.

\subsection{Key Derivation Function}
The key derivation function (\textbf{KDF}) is a cryptography function designed to derive one or more keys from a given parameter.  The main objective of KDF is to stretch keys to achieve a suitable length or convert keys into a required format. KDF usually take four different inputs: a random seed, a length, a salt s and context c. The security of KDF is captured from \cite{krawczyk2010cryptographic}. The advantage of any adversary A in probabilistic polynomial time to break the KDF security is defined as $Adv^{KDF}_{\mathcal{A}}$. 

\begin{center}

$Adv^{KDF}_{\mathcal{A}} = 2 * |Pr[(b=b^{'})- \frac{1}{2} ]|$.

\end{center}

For a sufficiently negligible value $\epsilon$, $Adv^{KDF}_{\mathcal{A}} < \epsilon$.

\section{System and Adversary Model}
\label{sec:sys_model}
This section briefly describes our proposed system and adversary model. In the system model, we show the different entities and their roles in the proposed scheme. In the adversary model, we consider possible potential attackers and their abilities against our proposed system. We first start with the system model, and then we introduce the adversary model.

\subsection{System Model}
As illustrated in Figure \ref{fig:sys_model}, our proposed system adheres to the conventional GAN structure, comprising two primary components of generator: an encoder and a decoder. Throughout the training phase, both the generator and discriminator are active. However, during deployment and inference, only the generator is operational. We've partitioned the generator into two segments to accommodate the specific requirements of split learning. One segment is located on the user end, while the other is hosted on the server side. On the other hand, due to its unique architecture, the discriminator solely finds its place on the server side. This strategic division bolsters data privacy and alleviates computational burdens on the user's end, thereby economizing device deployment. Our system's operation is split into two phases. The first phase leverages a pre-trained model for inference while simultaneously undergoing adaptive training. This phase is predominant at the system's inception when there's a dearth of user-specific data to facilitate comprehensive model training. Given the considerable disparities in household electricity consumption patterns and appliance variations among users, adaptive training becomes imperative. This ensures the model is tailored to individual users, guaranteeing optimal performance. The following phase is the stabilization phase. At this juncture, the model has largely been trained and is proficient in conducting detections. However, any unforeseen data alterations—perhaps due to personal adjustments or the introduction/removal of electrical appliances—can trigger the system to revert to the adaptive training phase. This automatic shift ensures that the model continuously refines its predictive accuracy in the face of changing data landscapes.

\begin{figure*}[t!]
\center{\includegraphics[scale=0.40,trim=0 0 0 0,clip]{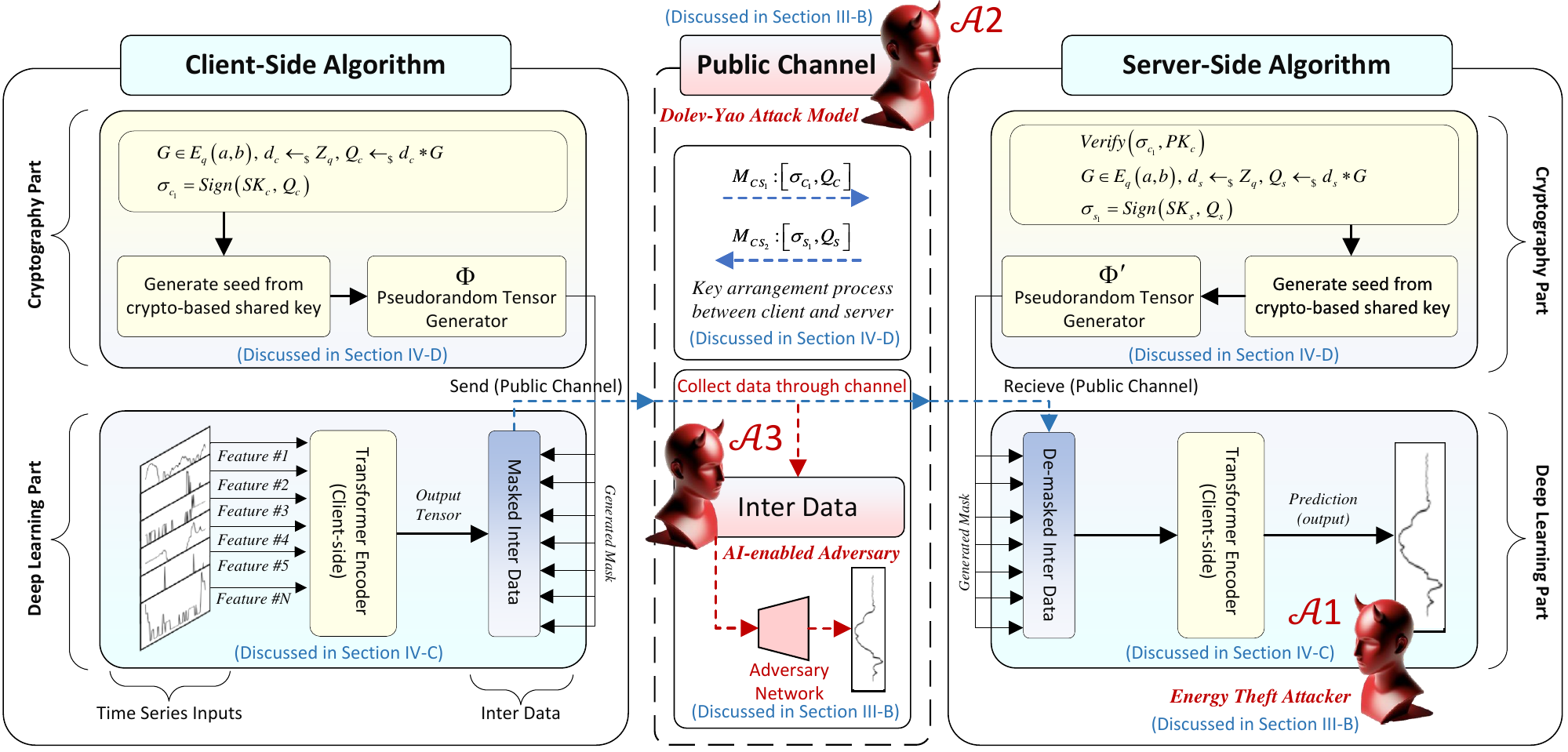}}
\caption{{Threat Model}}
\label{fig:adv_model} 
\vspace{-6mm}
\end{figure*}

\subsection{Adversary Model}
\label{sec:adversary_model}

In this paper, our primary concern focuses on both energy thefts and privacy protection threats. Privacy-based threats have become increasingly interested in collaborative machine learning models due to their widespread adoption in various applications. These threats can be launched by powerful adversaries who are equipped with artificial intelligence techniques. Hence, we consider AI-enabled adversarial attacks aimed at the user's privacy. This has led us to consider the following adversaries in our threat model:

\textbf{$\mathcal{A}_1$:} This type of adversary could be a malicious customer who may try to attempt to launch an energy theft attack. This adversary can modify meter readings using different attack scenarios. We consider three levels of attacks introduced in the threat models of \cite{jokar2015electricity} and \cite{alromih2022privacy}. Our model incorporates energy thefts that are launched by changing the meter readings of the \textbf{Consumer Smart Meter (CSM)}, where the adversary increases the meter readings for electricity theft. This indicates the following energy theft attack: $\mathcal{A}_1$ tries to reduce their consumption smart meter reading (CSM) by a percentage for a period of time T. This type of attack can cause a huge amount of financial loss in case of the insufficient detection. Our experiment in Section \ref{sec:exp_GAN-Transformer} evaluates how efficiently our model can detect energy theft compared with other state-of-art models.

\textbf{$\mathcal{A}_2$:} This type of adversary is an external attacker who may try to eavesdrop on the communication channel either physically or through cyber-attacks (Dolev-Yao Model). This enables the attacker to capture all or some messages transmitted through the public channel to try inferring individuals’ private data. This adversary is passive, compromising the privacy of the customers' data. This type of attacker can damage the user's privacy and money. Our experiment(s) and comprehensive security analysis prove that our proposed model can maintain the security and privacy of the message sent by the client and server.

\textbf{$\mathcal{A}_3$:} Here, we consider another type of adversary against the customers' privacy with more powerful capabilities. In conventional split learning, the model intermediate data (hidden layer feature) are often transmitted between client and server in plaintext form because of their large volume. In this regard, we consider the third type of adversary $\mathcal{A}_3$, which can use artificial intelligence (AI) to analyse the messages (intermediate data) sent between customers to extract private information. This AI-enabled adversary uses AI and advanced neural network structure to reconstruct the original raw data as discussed in \cite{salem2020updates}. This type of adversary can cause a very serious privacy issue. Our experiments in Section \ref{sec:exp_Complexity} and \ref{sec:exp_Privacy} evaluate how efficiently our proposed system can protect against this type of the adversary and how our proposed method is much faster than the traditional encryption methods for the same privacy level.

\section{Our Proposed Scheme}

In this section, we introduce our proposed GAN-Transformer. After that, we explain how our secure split learning protocol maintains the privacy issues in a split learning framework. Finally, we give the details of our proposed GAN-Transformer for energy theft detection in smart grids. Before diving into details, we give a high-level overview of each part.

\subsection{High Level Overview}
Now, we present a comprehensive overview (also shown in Figure \ref{fig:sys_model}) of our proposed work. As discussed in Section \ref{sec:sys_model}, our proposed system comprises three major components: algorithm for generative adversarial networks in split learning, secure split learning protocol, and GAN-Transformer. In this paper, we aim to provide a fully privacy-preserved system in anomaly detection based on split learning. These three components need to collaborate with each other to provide secure requirements. We propose the \emph{first} GAN-based transformer for energy theft detection combined with protocol-level protection to solve the electricity theft problem the smart grids, which perform better than the current literature. There is little research on the GAN-based model for split learning, and we proposed the first framework to fill the gap. Meanwhile, split learning is also vulnerable to deep-learning-based attacks such as reconstruction-based attacks; we propose the first protocol-based masking scheme against the AI-enable adversarial. We first introduce our GAN-based transformer for energy theft detection.

\subsection{GAN-Transformer}
Here, we give the details of our proposed model. Since the smart grid data is a time series data, so the model should be able to capture the time features. Our objective is to find the energy theft activities, and several approaches have been proposed in the literature, like autoencoder-based anomaly detectors \cite{an2015variational}, LSTM-based detectors \cite{malhotra2016lstm} and transformer-based detectors \cite{tuli2022tranad}. Meanwhile, the generative adversarial network has performed well in different areas, including anomaly detection. Based on that, we propose our GAN-based transformer for energy theft detection. Our GAN-based transformer mainly consists of two different structures, the generator and the discriminator. Our proposed generator has two different parts: the encoder and the decoder; we will discuss this part in the following subsection. We use a transformer encoder \cite{dosovitskiy2020image} with positional encoding to predict the total electricity usage of the client. Meanwhile, we use a transformer-based discriminator to force the generator to output more accurate data by providing adversary loss $\mathcal{L}_{Adv}$.

\subsection{Generative Adversarial Networks in Split Learning}

In this section, we introduce our GAN-based split learning framework. Due to the special structure of GAN, it is hard to implement split learning. Our proposed approach is to divide the generator into two different parts and keep the discriminator on the server side. By doing this, we can send the masked inter-data of the generator and encrypted data of total electricity usage. Since, inter-data is much bigger than the total electricity usage due to the tensor size and gradient, we use split learning to protect the user's privacy. Figure \ref{fig:sys_model} shows the details of the model structure. We split the transformer encoder into two parts and employ it on both the client side and the server side. Algorithm 1 and Algorithm 2 show the details of our proposed split learning framework combined with our proposed protocol.

\begin{algorithm}
\label{al:train}
\caption{Split-Learning Framework Training}
\begin{algorithmic}[1]
\REQUIRE Training data \( X \)
\REQUIRE Initialisation parameters: $\theta_{Enc},\ \theta_{Dec},\ \theta_{Dis}$

\STATE $Enc,\ Dec$ is the Encoder and Decoder of the Generator

\STATE $Dis$ is the Discriminator

\ENSURE Train all parts of the model

\FOR{each (epoch) in range (Epoch)}
    \FOR{each training sample \( x,\ y \) in \((X) \)}

        \STATE \textbf{Part 1: Client Side}
        \STATE // Forward pass in client-side

        \STATE \( T_{Mid} = Enc(x) \), $T_{Target} = y$

        \STATE Send message $M_{CS_2}$ to server

        \STATE \textbf{Part 2: Server Side}
        \STATE // Forward and backward pass in server-side

        \STATE \( \hat{x} = Dec(T_{Mid}) \)
        \STATE $\mathcal{L}_{rec} = \mathbb{E}_{x\sim px}||T_{Target} - \hat{x} ||_2$
        \STATE $\mathcal{L}_{adv} = \mathbb{E}_{x\sim px} || Dis(T_{Target}) - \mathbb{E}_{x\sim px}Dis(\hat{x})||_2$

        \STATE Compute generator and discriminator gradient:
        \STATE $\nabla \theta_{Gen} = (\lambda_{rec}*\mathcal{L}_{rec} + \lambda_{adv}*\mathcal{L}_{adv}).backward()$
        \STATE $\nabla \theta_{Dis} = \mathcal{L}_{adv}.backward()$
        
        \STATE Update decoder and discriminator weights:
        \STATE \( \theta_{Dec} = \theta_{Dec} - \eta \nabla \theta_{Gen} \)
        \STATE \( \theta_{Dis} = \theta_{Dis} - \eta \nabla \theta_{Dis} \)
        
        \STATE Send message $M_{SC_2}$ to client

        \STATE \textbf{Part 3: Client Side}
        \STATE // Backward pass in client-side
        \STATE \( \theta_{Enc} = \theta_{Enc} - \eta \nabla \theta_{Gen} \)

    \ENDFOR
\ENDFOR

\end{algorithmic}
\end{algorithm}

\begin{algorithm}
\label{al:test}
\caption{Split-Learning Framework Testing}
\begin{algorithmic}[1]
\REQUIRE Testing data \( X \)
\REQUIRE Load parameters: $\theta_{Enc},\ \theta_{Dec},\ \theta_{Dis}$

\STATE $Enc,\ Dec$ is the Encoder and Decoder of the Generator

\STATE $Dis$ is the Discriminator

\FOR{each (epoch) in range (Epoch)}
    \FOR{each testing sample \( x,\ y \) in \((X) \)}

        \STATE \textbf{Part 1: Client Side}
        \STATE // Forward pass in client-side

        \STATE \( T_{Mid} = Enc(x) \), $T_{Target} = y$

        \STATE Send message $M_{CS_2}$ to server

        \STATE \textbf{Part 2: Server Side}
        \STATE // Forward pass in server-side

        \STATE \( \hat{x} = Dec(T_{Mid}) \)

        \STATE $error = ||T_{Target} - \hat{x} ||_2$

        \IF{$error > threshold$}

            \STATE Report An Anomaly Behavior
        \ELSE
            \STATE Continue
        \ENDIF       
    \ENDFOR
\ENDFOR

\end{algorithmic}
\end{algorithm}

\subsection{Proposed Protocol Level Approach for Securing Split-Learning against AI Adversary} \label{sec:proposed_masking}

Here, we introduce our proposed secure split learning protocol against AI-enabled adversaries. Figure \ref{fig:authentication} shows the steps of the protocol. Our protocol mainly consists of two entities, named the user and the server. In the setup phase of the protocol, the client and server generate the secret and public keys and send them to a third-party trusted authority. After that, the client first generates its signature and sends it to the server. Upon receiving the message from the client, the server first verifies the signature and then generates the message for key exchange and verification. Each client who wants to send data to the server must verify the identification. The protocol can be summarized by following steps:

\begin{itemize}

\item[]\textbf{Step $\mathbf{1}$:} $\mathbf{M_{CS_1}}$: $ \{ \sigma_{c_1} \ Q_c \} $.

When a client tries communicating with the server, it first generates a secret key pair $(PK_c, SK_c) \leftarrow_\$ KGen(\lambda)$ for digital signature. After that, the client randomly generates $d_c \leftarrow_\$ Z_q$ and computes $Q_c = d_c * G$ for elliptic-curve Diffie–Hellman. Finally, the client generates the signature $\sigma_{c_1} = Sign(SK_c,\ Q_c)$ and sends message $M_{CS_1}$ to the server.

\item[]\textbf{Step $\mathbf{2}$:} $\mathbf{M_{SC_1}}$: $ \{ \sigma_{s_1} \ Q_s \} $.

After receiving the message, the server first verifies the client's signature. If the server receives a valid digital signature, it also generates a secret key pair $(PK_s, SK_s) \leftarrow_\$ KGen(\lambda)$ for the digital signature. After that, the server randomly generates $d_s \leftarrow_\$ Z_q$ and computes $Q_s = d_s * G$ for elliptic-curve Diffie–Hellman with signing the digital signature. Finally, the server derives the encryption key $k_{Enc}$ and mask key $k_{Mask}$ from the Defile-Hellman results $(Q_c * d_s)$ using KDF.

\item[]\textbf{Step $\mathbf{3}$:} $\mathbf{M_{CS_2}}$: $ \{ \sigma_{c_2} \ m_1,\ m_2 \} $.

\iffullversion
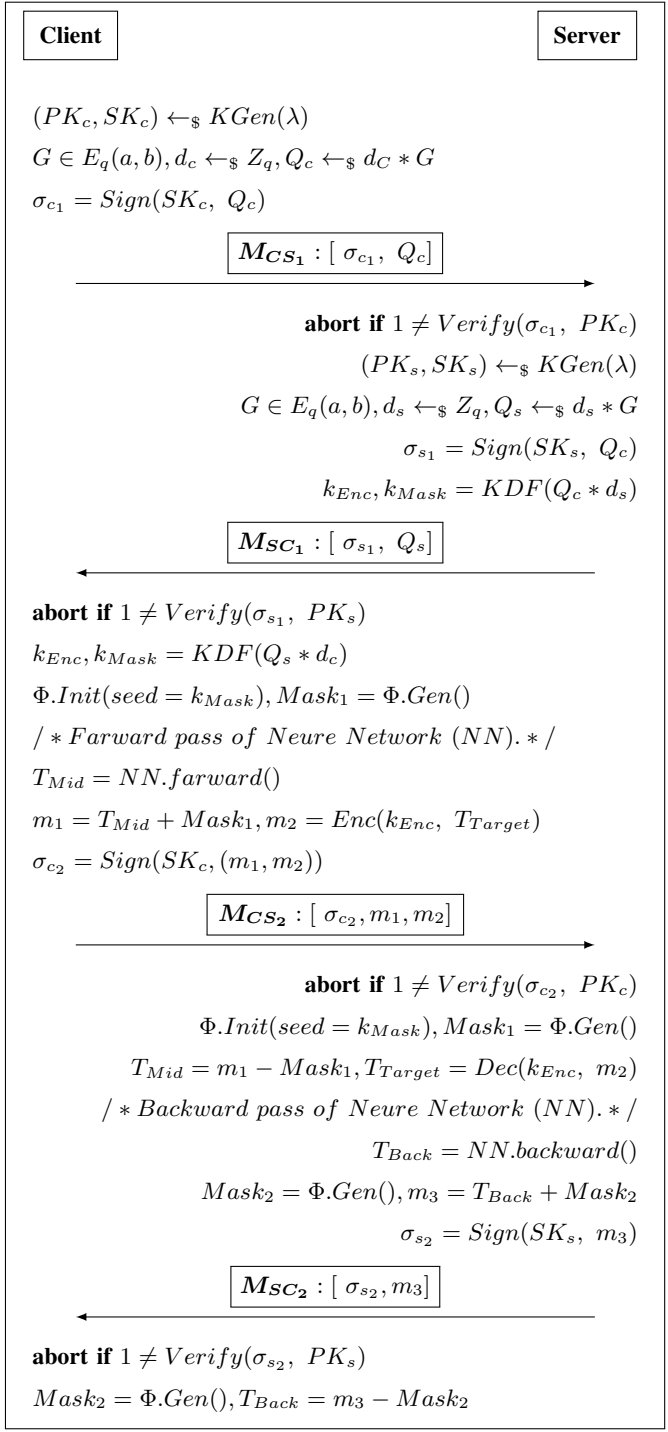
\begin{figure}[t!]
	\centering
	\begin{adjustbox}{max width=0.59\textwidth, max height=1\textheight}
		\fbox{
	\begin{tikzpicture}[yscale=-0.55,>=latex]
    \tikzstyle{every node}=[font=\small]
	\edef\InitX{-1}
	\edef\ArrowLeft{-0.3}
	\edef\ArrowCenter{6}
	\edef\ArrowRight{6.6}
	\edef\RespX{7.3}
	\edef\Y{0}
	
    \node [rectangle,draw,inner sep=6pt,right] at (\InitX,\Y) {\textbf{Client} };
    \node [rectangle,draw,inner sep=6pt] at (6.5,\Y) {\textbf{Server}};

    \NextLine[2]
    

    \ClientAction{$(PK_c,SK_c) \leftarrow_\$ KGen(\lambda)$}
    
    \NextLine
    \ClientAction{$ G \in E_q(a,b), d_c \leftarrow_\$ Z_q, Q_c \leftarrow_\$ d_C * G  $}

    \NextLine
    \ClientAction{$ \sigma_{c_1} = Sign(SK_c,\ Q_c) $}

    \NextLine[1.5]
    \ClientToServer{\framebox[1.1\width]{$\boldsymbol{M_{CS_1}}:[\ \sigma_{c_1},\ Q_c ]$}}{}

    \NextLine[0.5]
    \ServerAction{$ \textbf{abort if}\ 1 \neq Verify(\sigma_{c_1},\ PK_c)$}

    \NextLine
    \ServerAction{$(PK_s,SK_s) \leftarrow_\$ KGen(\lambda)$}

    \NextLine
    \ServerAction{$ G \in E_q(a,b), d_s \leftarrow_\$ Z_q, Q_s \leftarrow_\$ d_s * G  $}

    \NextLine
    \ServerAction{$ \sigma_{s_1} = Sign(SK_s,\ Q_c) $}

    \NextLine
    \ServerAction{$ k_{Enc},k_{Mask} = KDF(Q_c*d_s) $}

    \NextLine[1.5]
    \ServerToClient{\framebox[1.1\width]{$\boldsymbol{M_{SC_1}}:[\ \sigma_{s_1},\ Q_s ]$}}{}

    \NextLine[0.5]
    \ClientAction{$ \textbf{abort if}\ 1 \neq Verify(\sigma_{s_1},\ PK_s)$}

    \NextLine
    \ClientAction{$ k_{Enc},k_{Mask} = KDF(Q_s*d_c) $}

    \NextLine
    \ClientAction{$ \Phi.Init(seed = k_{Mask}), Mask_1 = \Phi.Gen() $}

    \NextLine
    \ClientAction{$/*Farward\ pass\ of\ Neure\ Network\ (NN).*/$}
    \NextLine
    \ClientAction{$ T_{Mid} = NN.farward() $}

    \NextLine
    \ClientAction{$ m_1 = T_{Mid} + Mask_1, m_2 = Enc(k_{Enc},\ T_{Target})$}

    \NextLine
    \ClientAction{$ \sigma_{c_2} = Sign(SK_c, (m_1, m_2)) $}

    \NextLine[1.5]
    \ClientToServer{\framebox[1.1\width]{$\boldsymbol{M_{CS_2}}:[\ \sigma_{c_2}, m_1, m_2 ]$}}{}

    \NextLine[0.5]
    \ServerAction{$ \textbf{abort if}\ 1 \neq Verify(\sigma_{c_2},\ PK_c)$}

    \NextLine
    \ServerAction{$ \Phi.Init(seed = k_{Mask}), Mask_1 = \Phi.Gen() $}

    \NextLine
    \ServerAction{$ T_{Mid} = m_1 - Mask_1, T_{Target} = Dec(k_{Enc},\ m_2)$}

    \NextLine
    \ServerAction{$/*Backward\ pass\ of\ Neure\ Network\ (NN).*/$}
    \NextLine
    \ServerAction{$ T_{Back} = NN.backward() $}

    \NextLine
    \ServerAction{$ Mask_2 = \Phi.Gen(), m_3 = T_{Back} + Mask_2 $}

    \NextLine
    \ServerAction{$ \sigma_{s_2} = Sign(SK_s,\ m_3) $}

    \NextLine[1.5]
    \ServerToClient{\framebox[1.1\width]{$\boldsymbol{M_{SC_2}}:[\ \sigma_{s_2}, m_3 ]$}}{}

    \NextLine[0.5]
    \ClientAction{$ \textbf{abort if}\ 1 \neq Verify(\sigma_{s_2},\ PK_s)$}

    \NextLine
    \ClientAction{$ Mask_2 = \Phi.Gen(), T_{Back} = m_3 - Mask_2  $}

	\end{tikzpicture}
}
	\end{adjustbox}	\caption{Proposed Protocol for Securing Split-Learning.}
	\label{fig:authentication}
\end{figure}
\else
\begin{figure}[htb]
	\centering
	\begin{adjustbox}{max width=1\textwidth, max height=1\textheight}
		\fbox{
	\begin{tikzpicture}[yscale=-0.55,>=latex]
    \tikzstyle{every node}=[font=\small]
	\edef\InitX{-1}
	\edef\ArrowLeft{-0.3}
	\edef\ArrowCenter{6}
	\edef\ArrowRight{6.6}
	\edef\RespX{7.3}
	\edef\Y{0}
	
    \node [rectangle,draw,inner sep=6pt,right] at (\InitX,\Y) {\textbf{Client} };
    \node [rectangle,draw,inner sep=6pt] at (6.5,\Y) {\textbf{Server}};

    \NextLine[2]
    

    \ClientAction{$(PK_c,SK_c) \leftarrow_\$ KGen(\lambda)$}
    
    \NextLine
    \ClientAction{$ G \in E_q(a,b), d_c \leftarrow_\$ Z_q, Q_c \leftarrow_\$ d_C * G  $}

    \NextLine
    \ClientAction{$ \sigma_{c_1} = Sign(SK_c,\ Q_c) $}

    \NextLine[1.5]
    \ClientToServer{\framebox[1.1\width]{$\boldsymbol{M_{CS_1}}:[\ \sigma_{c_1},\ Q_c ]$}}{}

    \NextLine[0.5]
    \ServerAction{$ \textbf{abort if}\ 1 \neq Verify(\sigma_{c_1},\ PK_c)$}

    \NextLine
    \ServerAction{$(PK_s,SK_s) \leftarrow_\$ KGen(\lambda)$}

    \NextLine
    \ServerAction{$ G \in E_q(a,b), d_s \leftarrow_\$ Z_q, Q_s \leftarrow_\$ d_s * G  $}

    \NextLine
    \ServerAction{$ \sigma_{s_1} = Sign(SK_s,\ Q_c) $}

    \NextLine
    \ServerAction{$ k_{Enc},k_{Mask} = KDF(Q_c*d_s) $}

    \NextLine[1.5]
    \ServerToClient{\framebox[1.1\width]{$\boldsymbol{M_{SC_1}}:[\ \sigma_{s_1},\ Q_s ]$}}{}

    \NextLine[0.5]
    \ClientAction{$ \textbf{abort if}\ 1 \neq Verify(\sigma_{s_1},\ PK_s)$}

    \NextLine
    \ClientAction{$ k_{Enc},k_{Mask} = KDF(Q_s*d_c) $}

    \NextLine
    \ClientAction{$ \Phi.Init(seed = k_{Mask}), Mask_1 = \Phi.Gen() $}

    \NextLine
    \ClientAction{$/*Farward\ pass\ of\ Neure\ Network\ (NN).*/$}
    \NextLine
    \ClientAction{$ T_{Mid} = NN.farward() $}

    \NextLine
    \ClientAction{$ m_1 = T_{Mid} + Mask_1, m_2 = Enc(k_{Enc},\ T_{Target})$}

    \NextLine
    \ClientAction{$ \sigma_{c_2} = Sign(SK_c, (m_1, m_2)) $}

    \NextLine[1.5]
    \ClientToServer{\framebox[1.1\width]{$\boldsymbol{M_{CS_2}}:[\ \sigma_{c_2}, m_1, m_2 ]$}}{}

    \NextLine[0.5]
    \ServerAction{$ \textbf{abort if}\ 1 \neq Verify(\sigma_{c_2},\ PK_c)$}

    \NextLine
    \ServerAction{$ \Phi.Init(seed = k_{Mask}), Mask_1 = \Phi.Gen() $}

    \NextLine
    \ServerAction{$ T_{Mid} = m_1 - Mask_1, T_{Target} = Dec(k_{Enc},\ m_2)$}

    \NextLine
    \ServerAction{$/*Backward\ pass\ of\ Neure\ Network\ (NN).*/$}
    \NextLine
    \ServerAction{$ T_{Back} = NN.backward() $}

    \NextLine
    \ServerAction{$ Mask_2 = \Phi.Gen(), m_3 = T_{Back} + Mask_2 $}

    \NextLine
    \ServerAction{$ \sigma_{s_2} = Sign(SK_s,\ m_3) $}

    \NextLine[1.5]
    \ServerToClient{\framebox[1.1\width]{$\boldsymbol{M_{SC_2}}:[\ \sigma_{s_2}, m_3 ]$}}{}

    \NextLine[0.5]
    \ClientAction{$ \textbf{abort if}\ 1 \neq Verify(\sigma_{s_2},\ PK_s)$}

    \NextLine
    \ClientAction{$ Mask_2 = \Phi.Gen(), T_{Back} = m_3 - Mask_2  $}

	\end{tikzpicture}
}
	\end{adjustbox}
	\caption{Protocol }
	\label{fig:authentication}
\end{figure}
\fi

Upon receiving the message from the server, the client first verifies the server's digital signature. After verifying the signature, the client derives the encryption key $k_{Enc}$ and mask key $k_{Mask}$ from the Diffie-Hellman results $(Q_s * d_c)$ using Key Derivation Function (KDF). Next, the client initializes the pseudorandom generator $\phi$ using mask key $k_{Mask}$ as seed and generates $Mask_1$ to transform $T_{Mid}$. $T_{Mid}$0 is the middle tensor of the neural network, and the client needs to forward pass the input data and then get the $T_{Mid}$. After getting the $T_{Mid}$, the client first adds $Mask_1$ to the $T_{Mid}$ by doing $m_1 = T_{Mid} + Mask_1$ and encrypts the target value to $m_2 = Enc(k_{Enc},\ T_{Target})$. Finally, the client sends messages $m_1$ and $m_2$ with a signature.

\item[]\textbf{Step $\mathbf{4}$:} $\mathbf{M_{SC_2}}$: $ \{ \sigma_{s_2} \ m_3 \} $.

After the server receives the client's message, it first verifies the signature. If the server receives a valid signature, it also initializes the pseudorandom generator $\phi$ using the same mask key $k_{Mask}$ as seed and generates $Mask_1$ for the de-masking of $T_{Mid}$. After that, the server decrypts message $T_{Target} = Dec(k_{Enc},\ m_2)$. The server needs to perform the backpropagation of the neural network and get the gradient $T_{Back}$ for training. Then, the server generates another mask $Mask_2$ for the gradient and sends message $m_3 = T_{Back} + Mask_2$ with signature.

\item[]\textbf{Step $\mathbf{5}$:} Client update.

Upon receiving the message from the server, the client first verifies the server's digital signature. After verifying the signature, the client uses the pseudorandom generator $\phi$ to generate $Mask_2$ for the de-masking of $T_{Back} = m_3 - Mask_2$. Finally, the client updates the model and finalizes the backward pass.

\end{itemize}

\begin{tcolorbox}
\textbf{Remark:} Here, one might question the necessity of a new protocol. Are there any existing two-party protocols (such as TLS) that could be applied directly here? It is important to acknowledge that, to begin with, there is no such protocol that can be explicitly applied in this context to address the fundamental issue of split learning efficiently. Since communication channels facilitate the transmission of substantial volumes of data, whereas the current cryptographic protocol (such as TLS) relies on end-to-end encryption to secure intermediate data, which demands considerable computational resources and time; in contrast,  when combined with the protocol-level approach, our proposed masking method does not require any end-to-end encryption and can be employed to defend against AI-enabled adversaries with enhanced efficiency.
\end{tcolorbox}

\section{Formal Proof for Protocol(w.r.t  Dolev-Yao Attacker, i.e., $\mathcal{A}_2$ )}

In this section, we provide the formal security for our proposed protocol. Due to page limitations, we provide all our security frameworks and the last theorem here. We divided our formal proof into five different sections: mutual authentication security (MA-security), unlinkability, key indistinguishability, AI Security and system security. The full proof of security is provided in the Appendix. We first begin with the security frameworks.

\subsection{Security Frameworks}

In this section, we introduce the security frameworks we used in our formal proof. Our protocol aims for secure key exchange and authentication for the client and the server. Therefore, we consider mutual authentication, key indistinguishability, and unlinkability.
In line with the security model posited by Bellare-Rogaway \cite{bellare1993random}, our rigorous security proof is predicated upon a set of security games engaging a probabilistic polynomial time (PPT) adversary, denoted as $\mathcal{A}$, and a challenger, denoted as $\mathcal{C}$. The key of these games is that the adversary is deemed to win if it successfully compromises either mutual authentication or other security frameworks. Our protocol is considered secure in terms of all security properties only if no probabilistic polynomial time adversary $\mathcal{A}$ can win these games. We now proceed to elaborate on the specific definitions and constructs of the security games pertaining to mutual authentication.

\subsubsection{\textbf{Mutual Authentication}}

Here, we outline the main objective of $\mathcal{A}$ in the mutual authentication security game, adhering to the framework of the existentially unforgeable under-chosen-message attacks (EUF-CMA) security game. The notation $Exp^{EUF-CMA}_{\mathcal{A}}(\lambda)$ is used to signify the interaction between the challenger $\mathcal{C}$ and the adversary $\mathcal{A}$. Specifically, in this context, the EUF-CMA game is applied to authenticate a signature scheme. The formal definition of this game is as follows:

\begin{enumerate}

    \item The challenger $\mathcal{C}$ generates a public key and private key pair $(pk, sk)$ using elliptic curve parameter and gives the adversary $\mathcal{A}$ the public key $pk$.

    \item The adversary $\mathcal{A}$ is allowed use adaptive chosen messages $m_1, m_2,...,m_q$ for some $q \in N$ to query the challenger $\mathcal{C}$.

    \item After the adversary $\mathcal{A}$ asks all its queries, $\mathcal{A}$ outputs a message pair $(m, \sigma)$

\end{enumerate}

In the formally defined game above, $\sigma$ is a forged signature produced by adversary $\mathcal{A}$. We define EUF-CMA security after we introduce the adversary's queries.

\textbf{Adversary Queries.} We define all queries as the adversary $\mathcal{A}$'s behaviour during the EUF-CMA security game $Exp^{EUF-CMA}_{\mathcal{A}}(\lambda)$:

\begin{itemize}
    \item $Create(i)$: allow $\mathcal{C}$ to initialize a public key and private key pair $(pk_i, sk_i)$ using elliptic curve parameter and give the public key $pk_i$ to the adversary $\mathcal{A}$.

    \item $Send(m, i)$: allow $\mathcal{A}$ to send a message $m$ to the challenger $\mathcal{C}$ and return a produced signature $\sigma_i = Sig(sk_i, m_i)$.

    \item $Corrpt(i) \rightarrow sk_i$: allow challenger $\mathcal{C}$ to leak its key $sk_i$.

    \item $StateReveal(i) \rightarrow P$: allow adversary $\mathcal{A}$ to reveal the internal state of $(pk_i, sk_i)$.
\end{itemize}

\noindent{\textbf{Definition 1} (\textbf{EUF-CMA} security)}: A signature scheme with functions: $(KGen, Sig, Verify)$ with security parameter $\lambda$. For a given cleanness predicate, and a probabilistic polynomial time (PPT) adversary $\mathcal{A}$, we define the advantage of $\mathcal{A}$ in EUF-CMA security game to be:

\begin{center}
    $Adv^{EUF-CMA}_{\mathcal{A}}(\lambda) = Pr( (m, \sigma) \leftarrow_\$ \mathcal{A}:Vf(pk, m, \sigma) = 1 \wedge m\notin \{ m1, m2,...,m_q \})$.
\end{center}

We say that the signature scheme holds EUF-CMA security if for all PPT $\mathcal{A}$, $Adv^{EUF-CMA}_{\mathcal{A}}(\lambda)$ is a negligible value in a parameter $\lambda$.

\subsubsection{\textbf{Key Indistinguishability}}

Here we outline the main objective of $\mathcal{A}$ in the key indistinguishability security game, referred to as KIND, and outline the types of queries $\mathcal{A}$ is permitted to make. $Exp^{KIND}_{\mathcal{A}, \Pi} (\lambda)$ denotes the game involving a challenger $\mathcal{C}$ and an adversary $\mathcal{A}$, with respect to the protocol $\Pi$. Here, $\lambda$ represents the security parameter associated with the protocol and we use elliptic curve digital signature algorithm as an example. The focus of this game is to evaluate the indistinguishability of keys generated by protocol $\Pi$. The formal structure and rules of this game are as follows:

\begin{enumerate}
    \item The challenger $\mathcal{C}$ randomly generates a $sk \rightarrow \{0, 1\}^n$ and computes $P = sk*G$.

    \item The adversary $\mathcal{A}$ chooses two messages $m_0 \rightarrow \{0, 1\}^n, m_1 \rightarrow \{0, 1\}^n$ and sends them to the challenger $\mathcal{C}$.

    \item The challenger computes:
    \begin{center}
        $b \leftarrow_\$ \{1, 0\}, \psi \leftarrow_\$ (Q = m_b * P = m_b * sk * G)$.
    \end{center}

    \item The adversary $\mathcal{A}$ outputs a guess $ \hat{b} \in \{0, 1\}$.

\end{enumerate}

The following is the formally defined game between an adversary $\mathcal{A}$ and a challenger $\mathcal{C}$. We first present the adversary queries.

\textbf{Adversary Queries.} We define all queries as the adversary $\mathcal{A}$'s behaviour during the key indistinguishability security game $Exp^{KIND}_\mathcal{A} (\lambda)$:

\begin{itemize}
    \item $Create(i)$: allow the challenger $\mathcal{C}$ to initialize a new secret key $sk_i$ and compute $P_i = sk_i*G$. 

    \item $Send(m, i)$: allow the adversarial $\mathcal{A}$ to send messages $m$ to the challenger $\mathcal{C}$ and return a produced message $m^{'}$.

    \item $Corrpt(i) \rightarrow sk_i$: allow the challenger $\mathcal{C}$ to leak its key $sk_i$.

    \item $StateReveal(i) \rightarrow P$: allow the challenger $\mathcal{C}$ to reveal the computed point $P_i = sk_i*G$.
\end{itemize}

\noindent{\textbf{Definition 2} (\textbf{Key Indistinguishability})}: Let $\Pi$ be a key exchange protocol. For a given cleanness predicate clean, and a probabilistic polynomial time (PPT) adversary $\mathcal{A}$, we define the advantage of $\mathcal{A}$ in the key indistinguishability game to be:

\begin{center}
    $Adv^{KIND}_{\mathcal{A}, \Pi}(\lambda) = |Pr(\hat{b}=b) - 1/2|$.
\end{center}

We say that $\Pi$ is KIND-secure if for all PPT $\mathcal{A}$, $Adv^{KIND}_{\mathcal{A}, \Pi}(\lambda)$ is a negligible value in a parameter $\lambda$.

\subsubsection{\textbf{Unlinkability}}

Here we describe the the primary aim of $\mathcal{A}$ in the unlinkability security game, as well as the queries that $\mathcal{A}$ can access. The game, denoted as $Exp^{Unlink}_{\mathcal{A},\Pi}(\lambda)$, involves a contest between a challenger $\mathcal{C}$ and an adversary $\mathcal{A}$ with respect to the protocol $\Pi$. The parameter $\lambda$ signifies the security parameter. The unlinkability security game, in this case, is formally described as follows:

\begin{enumerate}

    \item The challenger $\mathcal{C}$ initializes for each party and samples a random bit $b \leftarrow \{ 0,1 \}$.

    \item After that, the challenger $\mathcal{C}$ interacts with the adversary $\mathcal{A}$ via the adversary queries.

    \item Adversary $\mathcal{A}$ output a guess bit $b^{'}$.

\end{enumerate}

We now present the formally defined game between an adversary $\mathcal{A}$ and a challenger $\mathcal{C}$. We first introduce the adversary queries.

\textbf{Adversary Queries.} We define all queries as the adversary $\mathcal{A}$'s behaviour during the Unlinkability security game $Exp^{Unlink}_{\mathcal{A},\Pi}(\lambda)$ as follows. In addition to the $Creat,\ Corrupt$ $ Send\ and\ StateReveal$ queries listed above, we define two different queries:

\begin{itemize}

    \item Test($s,i,s^{'},i^{'}$) $\leftarrow \ m$: allow the adversary $\mathcal{A}$ to begin a new session $\pi_b$, where $b$ is sampled by a challenger $\mathcal{C}$. Here, $\pi_0 = \pi^s_i$ or $\pi_1 = \pi^{s^{'}}_{i^{'}}$ and they are both clean.

    \item SendTest($m$) $\leftarrow$ $m$: allow adversary $\mathcal{A}$ to send the message $m$ to session $\pi_b$.

\end{itemize}

\noindent{\textbf{Definition 3} (\textbf{Unlinkability})}: Let $\Pi$ be a key exchange protocol. For a given cleanness predicate clean, and a probabilistic polynomial time (PPT) adversary $\mathcal{A}$, we define the advantage of $\mathcal{A}$ in the unlinkability game to be:

\begin{center}
    $Adv^{Unlink}_{\mathcal{A}, \Pi}(\lambda) = |Pr(\hat{b}=b) - 1/2|$.
\end{center}

We say that $\Pi$ is Unlinkability-secure if for all PPT $\mathcal{A}$, $Adv^{Unlink}_{\mathcal{A}, \Pi}(\lambda)$ is a negligible value in a parameter $\lambda$.

\noindent{\textbf{Definition 4} (\textbf{AI Security})}: Let $\Pi$ be a deep learning model. For a given cleanness predicate clean, and a probabilistic polynomial time (PPT) adversary $\mathcal{A}$, we define the advantage of $\mathcal{A}$ in the AI game to be:

\begin{center}
    $Adv^{AI}_{\mathcal{A}, \Pi}(\lambda) = |Pr(\hat{b}=b) - 1/2|$.
\end{center}

We say that $\Pi$ is AI-secure if for all PPT $\mathcal{A}$, $Adv^{AI}_{\mathcal{A}, \Pi}(\lambda)$ is a negligible value in a parameter $\lambda$.


\textbf{Theorem 5:} Our protocol system holds the full security for any PPT time $\mathcal{A}$ under MA-security, key indistinguishability, unlinkability and AI security. The advantage of the adversary $\mathcal{A}$ in the full security games is $Adv^{Full}_{\mathcal{A}, \Pi}(\lambda)$.

\textbf{Proof:} First, recall that in order to break the full security of the system, the adversary $\mathcal{A}$ must break the MA-security, key indistinguishability, unlinkability and AI security. We first give the original attack game:

\textbf{Game 5.0: }This is the original attack game. We claim that: 

\begin{center}

    $Adv^{Full}_{\mathcal{A}, \Pi}(\lambda) < Adv_{G_{5.0}}$.
    
\end{center}

\textbf{Game 5.1: }In this game, the abort event is that the adversary $\mathcal{A}$ breaks any of the security properties. Thus, the advantage that $\mathcal{A}$ wins is bonded by the advantage of breaking any of the security properties:

\begin{center}
    \begin{adjustbox}{max width=0.49\textwidth, max height=1\textheight}
        $Adv_{G_{5.0}} < Adv_{G_{5.1}} + ANY (Adv^{MA}_{\mathcal{A}}, Adv^{KIND}_{\mathcal{A}},  Adv^{Unlink}_{\mathcal{A}}, Adv^{AI}_{\mathcal{A}})$.
    \end{adjustbox}
\end{center}

\textbf{Game 5.2: }In this game, the advantage of breaking any of the security properties is negligible based on Theorem 1 to Theorem 4. Thus, the advantage of $\mathcal{A}$ winning the full system security game is negligible: 

\begin{center}

    $Adv_{G_{5.1}} = 0$.

\end{center}

\section{Discussion}
\label{sec:dis}

In this section, we explain how we build the experiments with different attack models and how we measure the performance of our proposed model. We start with the experiment's basic setup to introduce the experiment platform and metrics we use and also provide the exploratory data analysis. After that, we introduce the anomaly detection part. Then, we show the baseline encryption and mask methods for comparison with the proposed method. Last, we show the superiority of the proposed method based on extensive experiments. All code and pre-trained models can be found through this link \footnote{Code and pre-trained model can be found from here: \url{https://tinyurl.com/3se362rc}}.

\subsection{Implementation Details}
Here, we present the specific details of our implementation of the proposed scheme. Subsequently, we demonstrate the computational expenses associated with client-side and server-side energy consumption. We first describe the details of the testbed.

\subsubsection{Implementation Setup}
\begin{figure}[h]
\center{\includegraphics[scale=0.2,trim=0 0 0 0,clip]{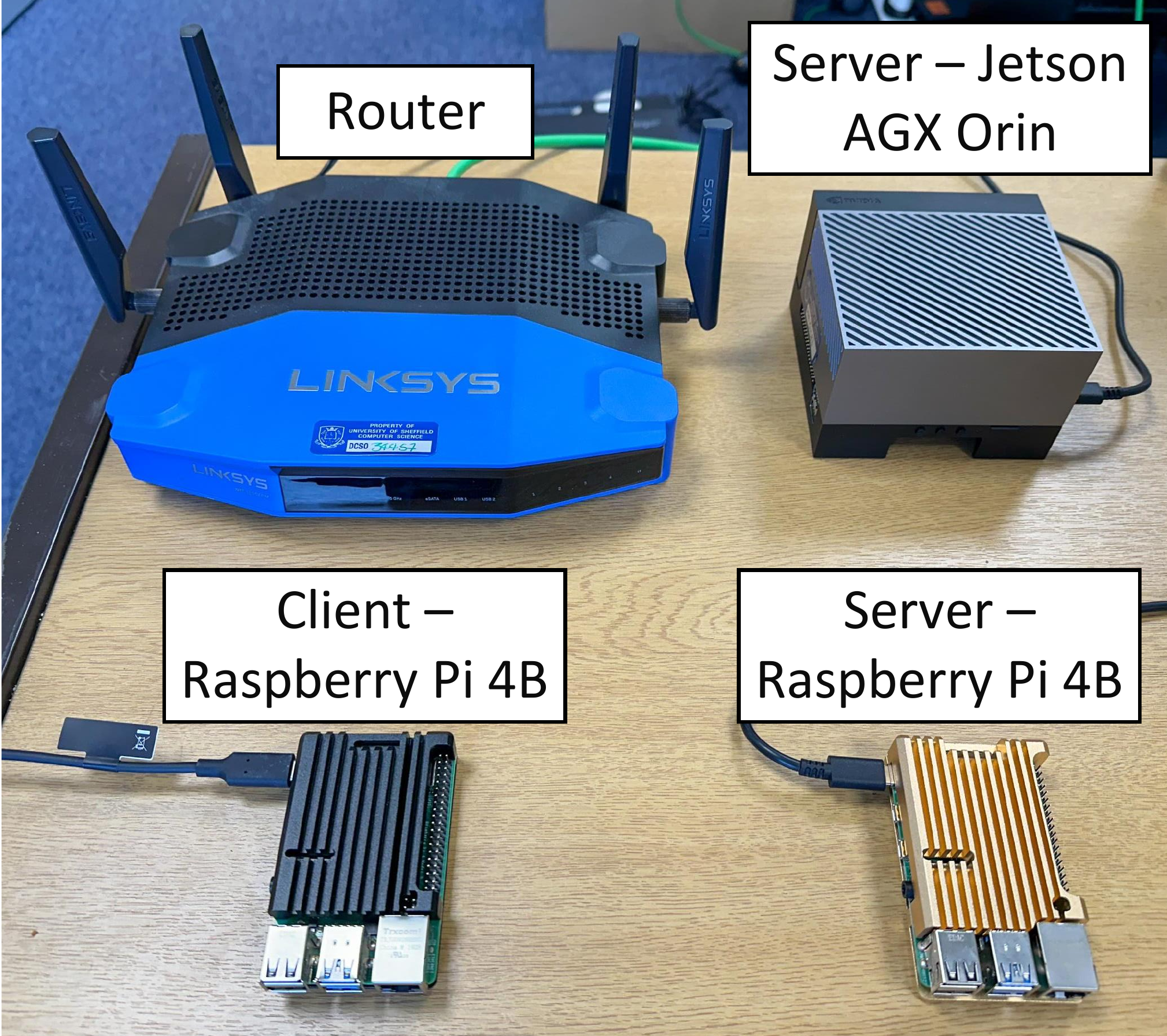}}
\caption{{Experiment Platform}}
\label{fig:Testbed} 

\end{figure}

We aimed to create a reliable and realistic experimental setup to evaluate the performance of our proposed scheme. To achieve the whole split-learning system, we involved two primary devices, namely the server and the client. For a more realistic test environment (the server has more computational resources), it was implemented on a Jetson AGX Orin equipped with a 12-core Arm CPU and 32GB RAM with an Ubuntu Jetson OS. Jetson also contains 2048 NVIDIA CUDA cores. For the execution of the client side, we employed the Raspberry Pi 4B, which is suitable for edge computing. All devices connect to the LINKSYS WRT3200ACM router. Figure \ref{fig:Testbed} shows the testbed.

Note that we implemented another server using Raspberry Pi; it is for the energy consumption experiment and provides a clear and fair comparison with the client-side device. The results of the energy consumption can be found in Section \ref{sec:energy}

\begin{table*}[ht]
\caption{Area Under Curve (AUC) Performance Comparison with related work}
\centering
\begin{tabular}{c c c c c c}
\toprule[1.5pt]
\multicolumn{6}{c}{{\textbf{Area Under Curve (AUC)}}} \\
\toprule[1.5pt]
\textbf{Adversarial Level} & \makebox[0.13\textwidth]{\textbf{AE (Conv) \cite{an2015variational}}} & \makebox[0.13\textwidth]{\textbf{GAnomaly \cite{akcay2018ganomaly}}} & \makebox[0.13\textwidth]{\textbf{LSTM \cite{hochreiter1997long}}} & \makebox[0.13\textwidth]{\textbf{Transformer \cite{vaswani2017attention}}} & \makebox[0.13\textwidth]{\textbf{Proposed Scheme}} \\
\cmidrule[1pt](lr){1-6}
10\% (more stealthy) & 0.498 & 0.492 & 0.657 & 0.643 & \textbf{0.690} \\
20\% & 0.497 & 0.486 & 0.707 & 0.784 & \textbf{0.817} \\
30\% & 0.497 & 0.483 & 0.952 & 0.967 & \textbf{0.970} \\
\midrule[1.5pt]
\end{tabular}
\label{tab:auc}
\end{table*}

\subsection{Experiment Setup}
\label{sec:exp_set}

Here, we introduce the experiment setup to introduce the metrics we use, and provide the dataset with the exploratory data analysis. We first introduce the dataset we use.

\subsubsection{Dataset with Exploratory Data Analysis}

\begin{figure}[h]
\center{\includegraphics[scale=0.23,trim=0 0 0 0,clip]{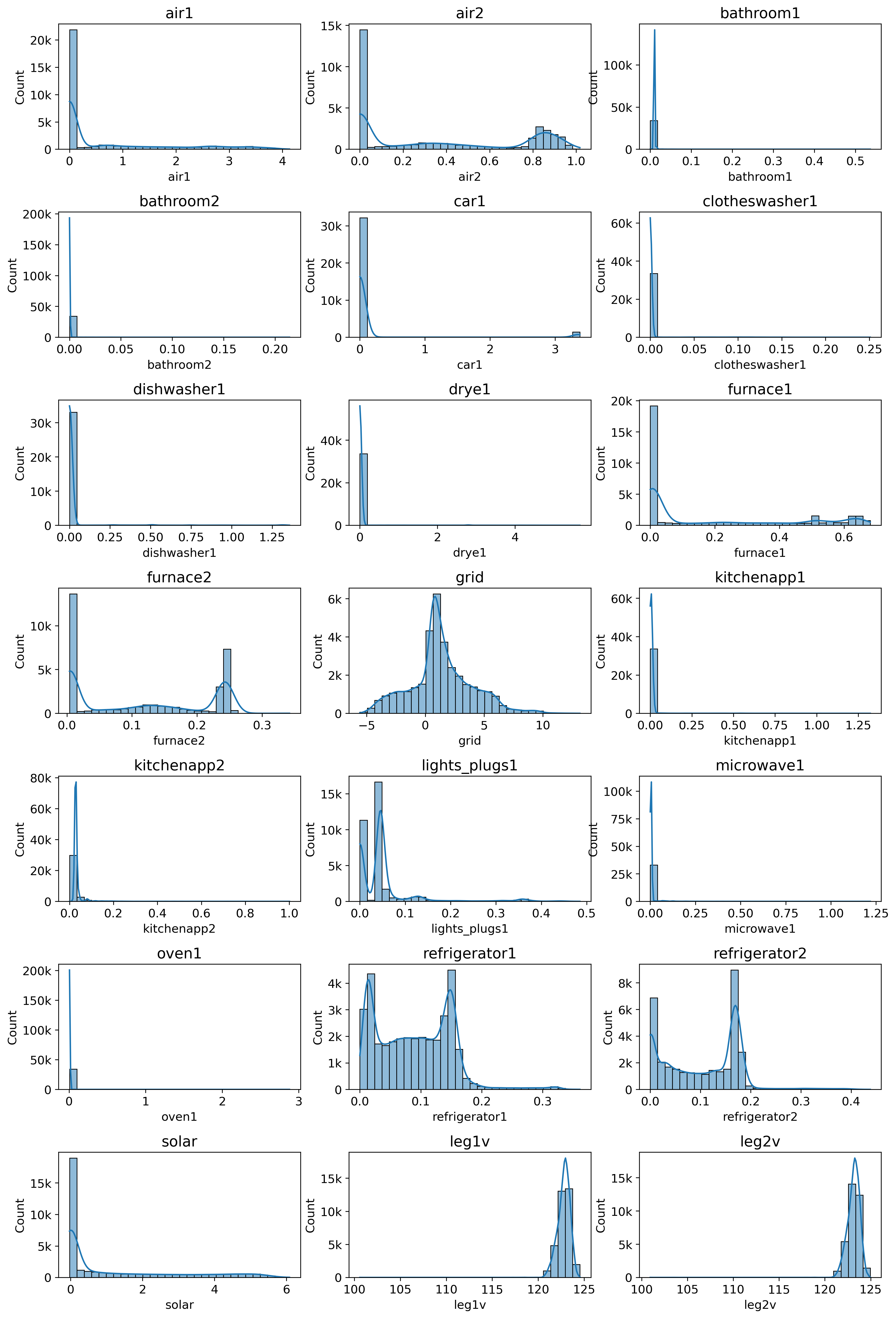}}
\caption{{Exploratory Data Analysis for Selected Features}}
\label{fig:EDA} 

\end{figure}

In the context of smart grid cybersecurity, particularly for challenges like energy theft detection, Exploratory Data Analysis (EDA) plays a crucial role. Utilizing the Pecan Street smart grid dataset, our analysis focuses on identifying patterns and anomalies in energy usage that could indicate fraudulent activities. This dataset encompasses a broad range of data, from traditional energy consumption metrics to advanced data from smart home devices including solar energy systems, electric vehicle charging stations, and smart meters.
Our EDA began with creating visualizations to analyze the distribution and interrelationships within the dataset. A primary method employed was histograms overlaid with Kernel Density Estimates (KDE), as shown in Figure \ref{fig:EDA}. This technique provides a comprehensive view of the data distribution, combining the direct frequency representation of the histogram with the continuous probability density function of the KDE. These visualizations are essential in spotting unusual patterns or outliers that may signal energy theft or vulnerabilities to AI adversarial attacks.

\begin{figure*}[htb]
\center{\includegraphics[scale=0.30,trim=0 0 0 20,clip]{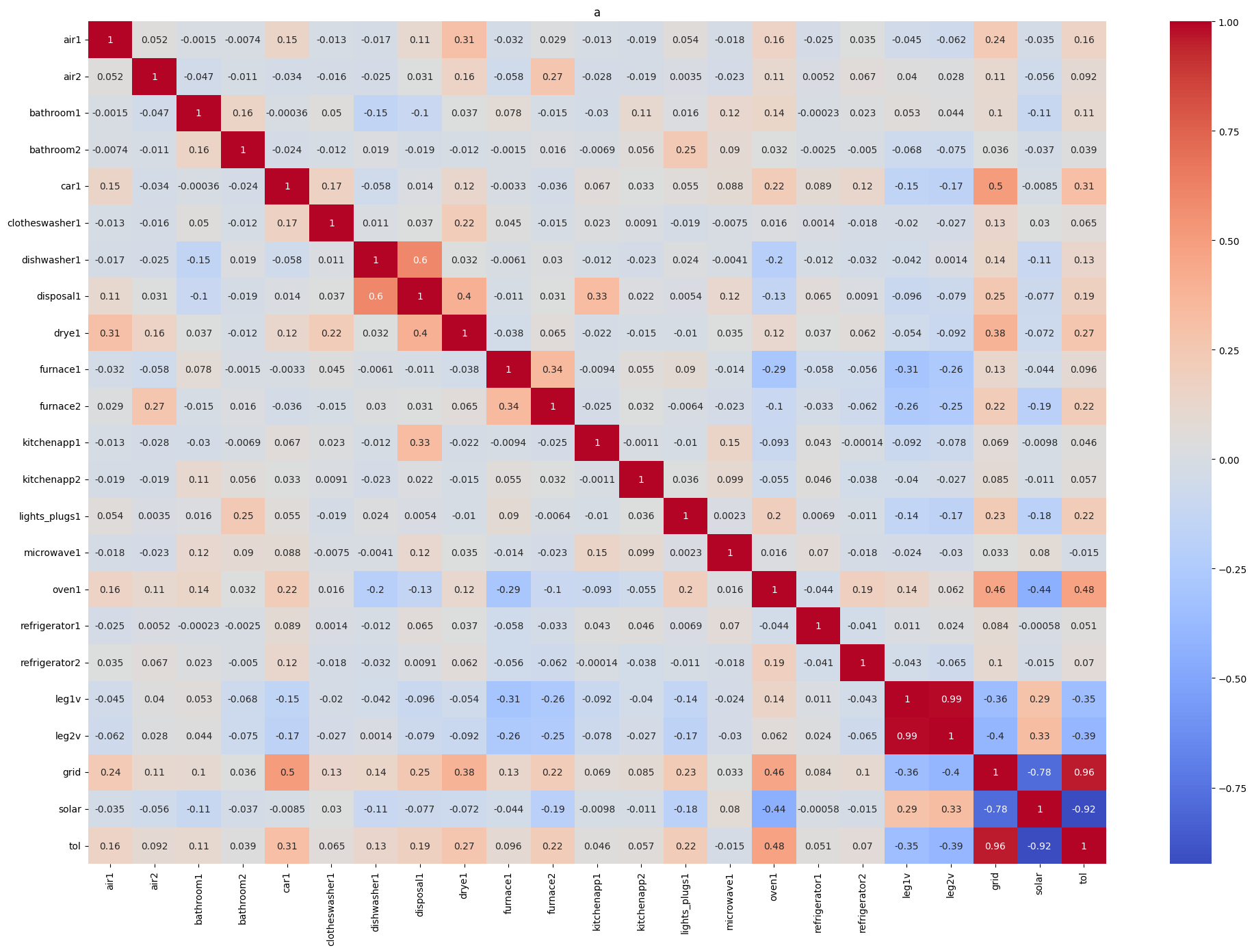}}
\caption{{Correlation Heatmap of Electricity Usage Features in the Pecan Street Smart Grid Dataset}}
\label{fig:EDA2} 
\end{figure*}

In addition, the feature correlation matrix, clearly illustrates how various energy consumption variables are interconnected. This understanding is key for effective feature engineering and normalization in our proposed models, which are used for detecting energy theft. The EDA process is not just a preliminary step; it is a critical component in unravelling the complexities of energy consumption behaviour and potential security threats in smart grids. The insights gained from this analysis are vital for developing more secure and efficient smart grid systems. They enable us to more accurately detect energy theft, thereby significantly contributing to the field of smart grid cybersecurity. As illustrated in Figure \ref{fig:EDA2}, these correlations are comprehensively analyzed, highlighting the intricate relationships between different variables. The correlation heatmap for the Pecan Street smart grid dataset unveils the interdependencies of electricity usage among various household appliances and energy sources. Strong correlations within appliance pairs, such as 'kitchenapp1' and 'kitchenapp2', are indicative of similar energy usage patterns. This may necessitate dimensionality reduction in further modelling to ensure model robustness. In stark contrast, notable negative correlations are observed between 'oven1' and 'solar' (-0.44) and 'drye1' and 'grid' (-0.38), indicating an inverse usage pattern potentially linked to solar energy availability.

Further analysis reveals a near-perfect correlation between 'leg1v' and 'leg2v' (0.99), suggesting redundant data capturing which requires careful feature selection to enhance model performance. 'Air1' and 'Drye1' display a moderate correlation (0.31), hinting at co-usage patterns possibly influenced by daily routines or climatic conditions. Additionally, 'disposal' correlates with both 'kitchenapp1' (0.33) and 'dishwasher1' (0.6), reflecting sequential tasks in kitchen activities. The total energy consumption ('grid') shows a moderate correlation with 'car1' (0.5), signifying the impact of electric vehicle charging on the energy profile. 'Lights\_plugs1' presents a lower correlation (0.23) with 'grid', suggesting varied influences on the total energy usage. These insights are pivotal, as they underscore the interconnected nature of the dataset's features and their collective influence on the energy management system. A noteworthy negative correlation between 'solar' (energy generation) and 'grid' (energy consumption) is essential for grid performance optimization and solar energy utilization enhancement. This analysis not only aids in understanding the data's inherent structure but also paves the way into energy theft detection, ultimately contributing to a more resilient and efficient energy grid.

\subsubsection{Experiment Metrics}
To better evaluate the performance of our proposed work, we report results using three different metrics. For experiment \ref{sec:exp_GAN-Transformer}, we use Area Under the Curve (AUC) to evaluate the overall performance of our proposed GAN-Transformer for energy theft detection. AUC excels in providing an aggregated measure of performance across all possible classification thresholds, thereby offering a comprehensive assessment of a model's discriminatory ability. It remains invariant under class distribution changes, making it a robust metric, particularly in imbalanced dataset scenarios. For experiment \ref{sec:exp_Complexity}, we measure the time usage for different encryption and decryption methods. For experiment \ref{sec:exp_Privacy}, we evaluate the power of the adversary decoder using the coefficient of determination $R^2$ \cite{chicco2021coefficient}. The Coefficient of Determination is a statistical measure that assesses the explanatory power of a predictive model, particularly in the context of regression analysis. $R^2$ offers a clear, numerical estimate of a model's predictive accuracy, facilitating objective comparisons between models. Its scale of 0 to 1 allows for intuitive interpretation, where 1 indicates perfect prediction and 0 implies no predictive capability. The $R^2$ is a standard metric for evaluating the goodness-of-fit in regression models, making it broadly applicable across various domains. In our experiment scenario where we train an attacker decoder adv-decoder, the aim is to assess its capability to extract or infer the user's raw data. In this case, the coefficient of determination can be employed to evaluate the adv-decoder’s effectiveness. Here, a high $R^2$ value indicates that the adv-decoder can accurately predict or replicate the outcomes it is designed to attack, signifying a strong attacking capability. Conversely, a low $R^2$ suggests a weaker attacking ability. 

\begin{figure*}[htb]
\center{\includegraphics[scale=0.55,trim=0 6 0 6,clip]{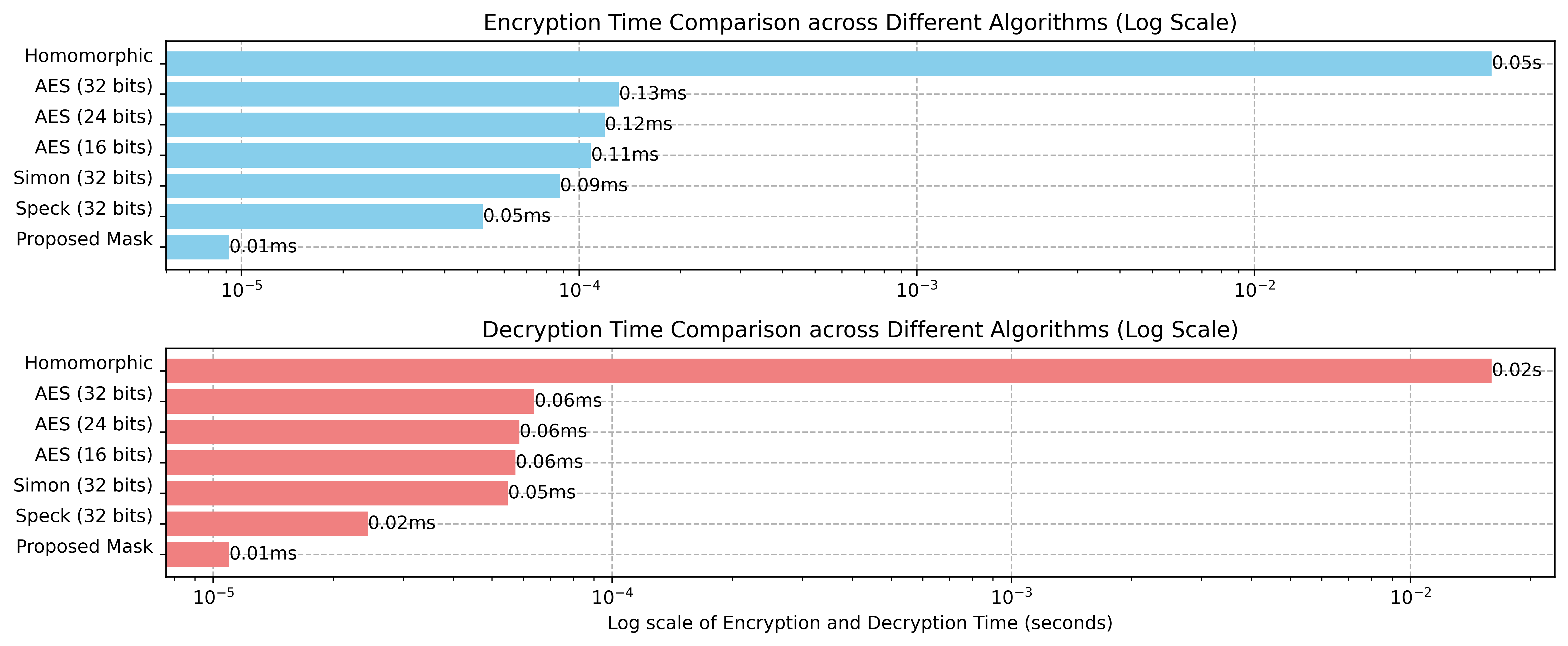}}
\caption{{Complexity Benchmark with the related encryption work}}
\label{fig:complexity_benchmark} 
\vspace{-6mm}
\end{figure*}

\subsection{GAN-Transformer Performance in Detecting Energy Theft Attacks (w.r.t  Energy Theft Attackers, i.e., $\mathcal{A}_1$)}
\label{sec:exp_GAN-Transformer}
This section details the evaluation of our proposed GAN-Transformer model's effectiveness in identifying energy theft attacks, specifically those executed by the adversary $\mathcal{A}_1$. We conducted a comprehensive set of experiments focusing on the AUC metric to gauge the model's detection capabilities. The evaluation spanned three distinct levels of adversarial energy theft, quantified at 0.1, 0.2, and 0.3, representing the proportion of stolen energy ranging from 10\% to 30\% (10\% is more stealthy). Beyond assessing the GAN-Transformer model's performance across these levels, we also benchmarked it against several leading deep learning models, including a conventional autoencoder (AE) \cite{an2015variational}, GAnomaly \cite{akcay2018ganomaly}, LSTM \cite{hochreiter1997long}, and Transformer \cite{vaswani2017attention}. The training phase of all models utilized authentic, non-compromised data devoid of any energy theft instances. Subsequently, the evaluation phase employed a balanced dataset comprising both genuine and compromised (theft) data points.
These experiments' summarised results are presented in Table \ref{tab:auc}.

The results indicate that our detector, even under low-level adversarial conditions (0.1), achieves a promising AUC of 0.690, demonstrating effective detection capabilities for minimally invasive energy theft scenarios. The detector's proficiency escalates with increasing adversarial levels, achieving an AUC of 0.817 at a 0.2 level and an impressive 0.970 at a 0.3 level. Significantly, our proposed GAN-Transformer model consistently outperformed other advanced deep learning models across all tested energy theft levels. It demonstrated at least a 5\% higher detection rate as compared to other high-performing models, such as the Transformer and LSTM (as shown in Table 2).

\begin{tcolorbox}
\textbf{Takeaway:} The results from our experiments with the GAN-Transformer model in detecting energy theft in smart grids reveal its exceptional effectiveness and versatility. Its consistently high AUC scores across varied adversarial levels, peaking at 0.970 in the most challenging 0.3 adversarial level, demonstrate an advanced ability to detect both subtle and significant theft scenarios. This is further exemplified by its impressive detection accuracy even at the lowest theft level (0.690 at 0.1 level), underscoring its sensitivity to minimal deviations, which is critical for early detection. This performance indicates a high degree of sensitivity and accuracy, which is critical in diverse and dynamic energy grid environments. Furthermore, the GAN-Transformer's superior performance over other advanced models like LSTM and standard Transformers highlights its unique strengths.
\end{tcolorbox}

\subsection{Complexity Analysis of the Proposed Mask Scheme with Other Encryption Mechanisms}
\label{sec:exp_Complexity}
This section evaluates the computational efficiency and complexity of our proposed masking scheme for privacy-preserving energy theft detection in comparison to traditional methods and other encryption techniques. We juxtapose our approach with widely used encryption mechanisms, including AES \cite{daemen1998block}, Simon \cite{beaulieu2013simon}, Speck \cite{beaulieu2013simon}, and homomorphic encryption \cite{gentry2009fully}, which are prevalent in privacy-preserving applications.
The primary source of complexity in these privacy-preserving methods, including ours, is centred around the processes of encryption (masking) and decryption (demasking). These steps are essential for securing data but often come with the trade-off of increased computational load.

Figures \ref{fig:complexity_benchmark} present all methods' encryption and decryption times. Our findings indicate that the proposed masking method is significantly more efficient than the alternatives. Notably, while homomorphic encryption is a common choice in deep learning applications for its security advantages, it is also one of the most complex encryption techniques. This complexity is primarily due to the intricate mathematical operations involved and the property of the homomorphic encryption \cite{munjal2023systematic}.

\begin{tcolorbox}
\textbf{Takeaway:} The proposed privacy-preserving energy theft detection scheme not only offers a comparable level of security to that of homomorphic and other advanced encryption methods but also stands out for its efficiency. This efficiency is crucial in AI-enabled attack scenarios, where the ability to rapidly process and protect large volumes of data is essential. Our approach demonstrates that achieving robust security without sacrificing computational efficiency is possible, making it a viable solution for real-time energy theft detection in smart grids.
\end{tcolorbox}

\begin{figure*}[htb]
\center{\includegraphics[scale=0.40,trim=0 0 0 0,clip]{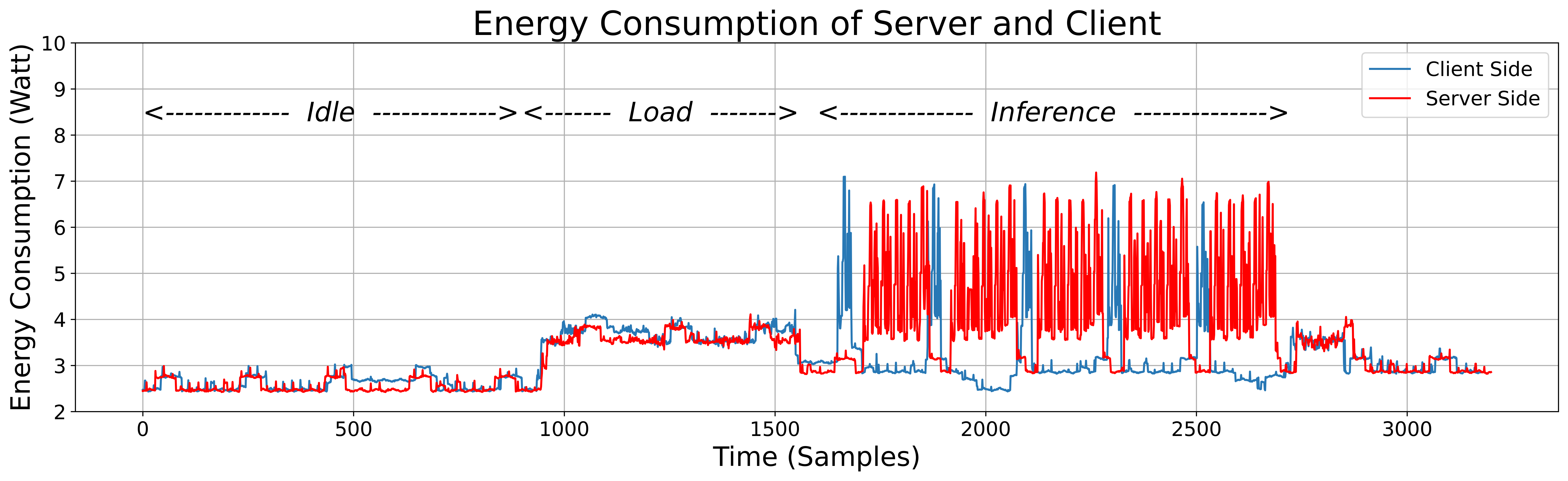}}
\caption{{Energy Consumption}}
\label{fig:energy} 
\end{figure*}

\subsection{Privacy Analysis of Our Proposed Scheme (w.r.t  AI-enabled Adversary, i.e., $\mathcal{A}_3$ )} 
\label{sec:exp_Privacy}

The proposed GAN-Transformer approach for privacy-preserving energy theft detection is potentially vulnerable to reconstruction attacks. These attacks aim to reconstruct the original energy consumption data from the inter-data sent between the system's components. If successful, reconstruction attacks can compromise the privacy of energy consumers. These attacks can be launched by an AI-enabled attacker, $\mathcal{A}_3$, which was explained earlier in Section \ref{sec:adversary_model}. Here, we explain how we managed this privacy issue by using our proposed masking protocol. As we discussed in Section \ref{sec:proposed_masking}, our proposed protocol has a shared key that is used to initialize a pseudonym generator to generate the mask for the tensor. This mask can hide and destroy the distribution of the inter-data. To conduct our privacy analysis, we assumed that an AI-enabled adversary intercepted some of the inter-data sent in the system and tried reconstructing the original data by training an adversarial decoder. 

\begin{figure}[htb]
\center{\includegraphics[scale=0.35,trim=5 0 0 0,clip]{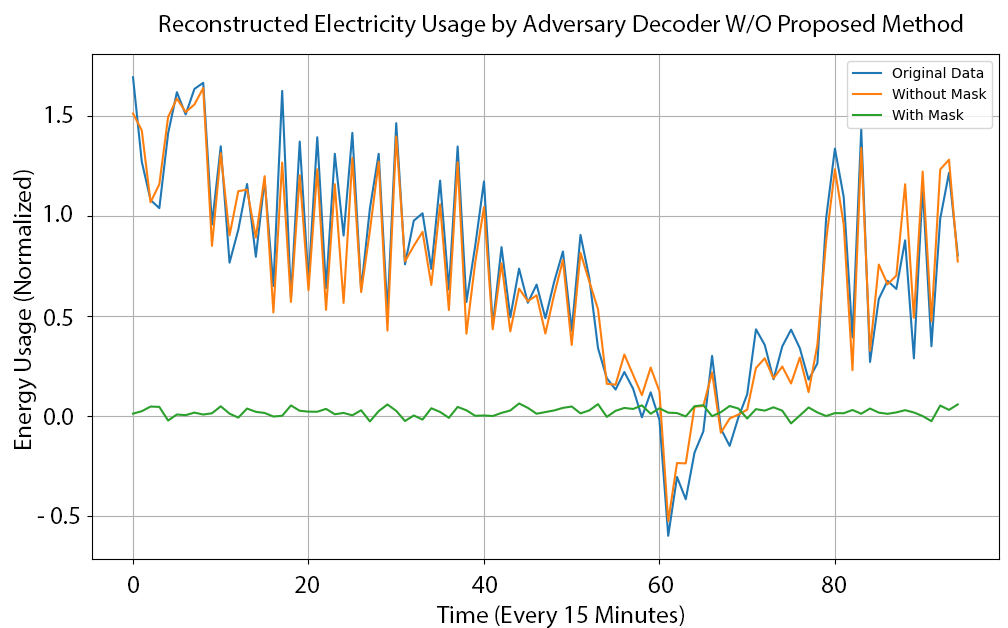}}
\caption{{Privacy Experiment results}}
\label{fig:privacy} 
\vspace{-3mm} 
\end{figure}

Figure \ref{fig:privacy} shows a sample result from our experiments which is a real-world energy usage where the x-axis signifies the index of samples taken at fifteen-minute intervals, and the y-axis depicts the normalized electrical energy data. The blue line indicates the real energy usage record, the orange line is the reconstructed output of the adversarial decoder without our proposed protocol, and the green line is the reconstructed data when the proposed masking is applied.  As we can see, this experiment shows how severe reconstruction attacks are when no security measures are taken. Looking at the orange line, we can see that the attacker can almost perfectly reconstruct the original data from only the inter-data. However, when the proposed masking approach is used, we can see that the adversarial decoder fails to reconstruct or extract any useful information from the data.

To better illustrate the performance of our proposed protocol, we measured how accurate it is to reconstruct the input back by using the coefficient of determination $R^2$ metric. The results of our experiments are shown in Table \ref{tab:r2}, where we show how accurate it is to reconstruct the original numbers by measuring the coefficient of determination $R^2$ metric. As we can see from the table, the $R^2$ is very high, indicating that the AI-enabled attacker can reconstruct the data with high accuracy without using our proposed masking protocol. In this case, the attacker can reconstruct the number from the inter-data. However,  with our proposed masking protocol, the values of $R^2$ are near zeros which indicates that the attacker cannot extract any useful information from the inter-data communication.

\begin{table}[ht]
\caption{R² values for different samples}
\begin{center}
\begin{tabular}{c c c c c c}
\toprule[1.5pt]
\multicolumn{6}{c}{{\textbf{Coefficient of determination (R²)}}} \\
\toprule[1.5pt]
\textbf{Sample Number} &  \makebox[0.04\textwidth] {\textbf{1}} &  \makebox[0.04\textwidth] {\textbf{2}} &  \makebox[0.04\textwidth] {\textbf{3}} &  \makebox[0.04\textwidth] {\textbf{4}} &  \makebox[0.04\textwidth] {\textbf{5}}  \\
\cmidrule[1pt](lr){1-6}

\textbf{no Mask} & 0.9746 & 0.9902 & 0.9773 & 0.9865 & 0.9814 \\
\textbf{with Mask} & \textbf{0.0095} & \textbf{0.0038} & \textbf{0.0036} & \textbf{0.0034} & \textbf{0.0011} \\

\midrule[1.5pt]

\end{tabular}
\label{tab:r2}
\end{center}

\begin{flushleft}
\begin{tablenotes}
        \footnotesize
        \item \textbf{Note:} Here we use efficient of determination $R^2$ to evaluate the power of an AI-enabled adversary $\mathcal{A}_3$. As we discussed in section \ref{sec:exp_set}, the lower number suggests a stronger defence ability.
\end{tablenotes}
\end{flushleft}

\end{table}

\begin{tcolorbox}
\textbf{Takeaway:} The analysis presented in this section underlines the significance of the proposed protocol-level approach for privacy preservation in energy theft detection, especially in the face of potential reconstruction attacks by an AI-enabled adversary, $A_3$. As we discussed in Section \ref{sec:adversary_model}, this application inversely correlates with defensive strength. A lower R2 value, indicative of a less capable attacker, implies a robust defence mechanism.
\end{tcolorbox}

\subsection{Energy Consumption} 
\label{sec:energy}

To better illustrate the effectiveness of our proposed scheme, we built an energy consumption experiment. Figure \ref{fig:energy} shows a time series plot of the Power (W: Watt) curve. We collected the power consumption in three stages, namely empty running (idle phase), loading phase, and inference phase. As we can see, the power of the inference phase is around 4-7W, which is quite low compared to the load and idle phases. Meanwhile, most of the computational overload is on the server side due to the split learning framework (refers to the red curve).

\begin{tcolorbox}
\textbf{Takeaway:} The experiment we provide above shows the entire energy consumption of all stages involved in our proposed scheme. All data collected from Raspberry shows that our scheme is suitable for edge computing environments like smart gird. The low energy consumption shows the effectiveness of our proposed GAN-enhanced techniques, which can achieve better performance with our increase in model complexity.
\end{tcolorbox}
\vspace{-3mm}

\section{Conclusion}

Securing the smart grids against energy theft and ensuring consumer privacy are essential for maintaining the resilience of these grids against unpredictable disruptions. 
Addressing these critical challenges, our research introduces an innovative GAN-Transformer-based split learning framework for energy theft detection in smart grids, leading to significant advancements in both privacy and efficiency in this domain. The proposed framework effectively leverages the transformer architecture's proficiency in handling long-range dependencies in energy consumption data, enabling precise detection of energy theft without compromising user privacy. Our innovative mask-based method marks a \emph{first} in the domain, effectively shielding against privacy leakage attacks during model training. Experimental results validate the framework's comparable accuracy to state-of-the-art energy theft detectors while providing significantly enhanced privacy protection.
Moreover, our complexity analysis confirms the superior efficiency of our masking scheme over traditional encryption methods, an essential attribute in AI-enabled attack scenarios where rapid data processing is crucial. This efficiency, combined with robust security, positions our approach as a viable solution for real-time energy theft detection in dynamic smart grid environments. The GAN-Transformer model's consistent outperformance of other advanced models across various adversarial levels underscores its unique strengths and potential as a cutting-edge tool in the cybersecurity domain. Therefore, our work not only contributes a novel approach to energy theft detection but also advances the field of privacy-preserving AI in smart grids.

\vspace{-6mm}

\bibliographystyle{IEEEtran}
\bibliography{main}

\end{document}